\documentclass[%
 aip,
 amsmath,amssymb,
 preprint,%
]{revtex4-1}

\usepackage{graphicx}
\usepackage{dcolumn}
\usepackage{bm}

\usepackage[utf8]{inputenc}
\usepackage[T1]{fontenc}
\usepackage{mathptmx}
\usepackage{etoolbox}

\usepackage{geometry}                		
\usepackage{graphicx}				

\usepackage{tikz}
\usetikzlibrary{shapes,arrows}
\usepackage{verbatim}
\usepackage{bm}

\usepackage[section]{placeins}
\usepackage{longtable}

\usepackage{lettrine}
\usepackage[retainorgcmds]{IEEEtrantools}
\usepackage{siunitx}       

\newcommand{\pa}{\partial}
\newcommand{\eqnref}[1]{(\ref{#1})}

\def\longrightleftharpoons{
  \mathop{
    \vcenter{
      \hbox{
      \ooalign{
        \raise1.5pt\hbox{$\longrightarrow\joinrel$}\crcr
	  \lower1.5pt\hbox{$\longleftarrow\joinrel$}
	  }
      }
    }
  }
}

\DeclareSymbolFont{sfletters}{OML}{cmbrm}{m}{it}
\DeclareMathSymbol{\stau}{\mathord}{sfletters}{"1C}

\makeatletter
\def\@email#1#2{%
 \endgroup
 \patchcmd{\titleblock@produce}
  {\frontmatter@RRAPformat}
  {\frontmatter@RRAPformat{\produce@RRAP{*#1\href{mailto:#2}{#2}}}\frontmatter@RRAPformat}
  {}{}
}%
\makeatother
\begin{document}

\preprint{AIP/123-QED}

\title[Non-equilibrium plasma generation via nano-second multi-mode laser pulses]{Non-equilibrium plasma generation via nano-second multi-mode laser pulses}

\author{Andrea Alberti}
\affiliation{The Center for Exascale Simulation of Plasma-coupled Combustion, University of Illinois at Urbana--Champaign, 1308 W.~Main St., Urbana, IL 61801, USA.}
\affiliation{Department of Aerospace Engineering, University of Illinois at Urbana--Champaign, 104 S.~Wright St., Urbana, IL 61801, USA.}

\author{Alessandro Munaf\`{o}}
\affiliation{The Center for Exascale Simulation of Plasma-coupled Combustion, University of Illinois at Urbana--Champaign, 1308 W.~Main St., Urbana, IL 61801, USA.}
\affiliation{Department of Aerospace Engineering, University of Illinois at Urbana--Champaign, 104 S.~Wright St., Urbana, IL 61801, USA.}

\author{Munetake Nishihara}
\affiliation{The Center for Exascale Simulation of Plasma-coupled Combustion, University of Illinois at Urbana--Champaign, 1308 W.~Main St., Urbana, IL 61801, USA.}

\author{Carlos Pantano}
\affiliation{The Center for Exascale Simulation of Plasma-coupled Combustion, University of Illinois at Urbana--Champaign, 1308 W.~Main St., Urbana, IL 61801, USA.}
\affiliation{Department of Mechanical Science \& Engineering, University of Illinois at Urbana--Champaign, 1206 W.~Green St., Urbana, IL 61801, USA.}

\author{Jonathan~B.~Freund}
\affiliation{The Center for Exascale Simulation of Plasma-coupled Combustion, University of Illinois at Urbana--Champaign, 1308 W.~Main St., Urbana, IL 61801, USA.}
\affiliation{Department of Aerospace Engineering, University of Illinois at Urbana--Champaign, 104 S.~Wright St., Urbana, IL 61801, USA.}
\affiliation{Department of Mechanical Science \& Engineering, University of Illinois at Urbana--Champaign, 1206 W.~Green St., Urbana, IL 61801, USA.}

\author{Marco Panesi}
\affiliation{The Center for Exascale Simulation of Plasma-coupled Combustion, University of Illinois at Urbana--Champaign, 1308 W.~Main St., Urbana, IL 61801, USA.}
\affiliation{Department of Aerospace Engineering, University of Illinois at Urbana--Champaign, 104 S.~Wright St., Urbana, IL 61801, USA.}

\email{mpanesi@illinois.edu.}

\date{\today}

\begin{abstract}
The formation and growth of plasma kernels generated via nano-second mode-beating laser pulses is investigated here via a non-equilibrium self-consistent computational model. The chemically reactive Navier--Stokes equations are used to describe the hydrodynamics, and non-equilibrium effects are taken into account with a two-temperature model. Inverse Bremsstrahlung and multiphoton ionization are included self-consistently in the model via a coupled solution of the plasma governing equations and the Radiative Transfer Equation (that describes the laser beam propagation and attenuation). The self-consistent approach (despite carrying additional challenges) minimizes the empiricism and it allows for a more accurate description, since it prevents both the utilization of artificial plasma seeds to trigger the breakdown and the implementation of tuning parameters to simulate the laser-energy deposition. The advantages of this approach are confirmed by the good agreement between the numerically predicted and the experimentally measured plasma boundary evolution and absorbed energy. This also holds true for the periodic plasma kernel structures that, as suggested by the experiments and confirmed by the simulations presented here, are connected to the modulating frequency.
\end{abstract}

\maketitle

\section{Introduction}
The first studies on optical breakdown of gases began in the 1960's, with the advent of powerful Q-switched lasers \cite{radziemski,morgan,ostrovskaya1974laser,morgan_1978,Raizer_1980}. Many applications have been extensively investigated, including laser-induced ignition \cite{massa2015fluid,massa2017plasma,starikovskaia2006plasma, mahamud2018dual,dumitrache2018numerical,dumitrache2016laser, butte2019properties,dumitrache2017control,sato2021numerical}, flow control \cite{limbach2015characterization,knight2008survey,inproceedingsKianvashrad,pournadali2019towards,starikovskiy2016trajectory, kianvashrad2019nonequilibrium,trefny2010supersonic,trefny2017dual, zheltovodov2007energy,adelgren2005control,alberti2019supersonic,alberti2020self,khamseh2021stereoscopic}, laser ablation \cite{panchenko2011pulsed}, and spectroscopy \cite{colonna2001modelling,musazzi2014libs}. 

Breakdown is observed when a laser beam of sufficient intensity is focused onto a gas (\emph{e.g.}, air) which is normally transparent to optical radiation \cite{radziemski,morgan,ostrovskaya1974laser,morgan_1978,Raizer_1980,harilal2017and}. Investigators were initially a bit puzzled by this finding as they were unable to justify breakdown on a theoretical basis. It was soon recognized that the disagreement between theory and experiments was caused by the reliance of the former on linear optics. As a matter of fact, for the large intensities attainable by Q-switched lasers, non-linear processes such as multiphoton ionization produce cold \emph{priming} electrons which, upon draining energy from the beam via inverse Bremsstrahlung, may trigger a cascade ionization process. This mechanism of plasma formation has been confirmed by both experiments and calculations. For a given gas, breakdown is not observed below a threshold intensity. The value of the former depends on several factors such as ambient conditions (\emph{e.g.}, pressure and temperature), beam wavelength and optics of the focusing system (\emph{e.g.}, focal length), and is also affected by the presence of impurities (\emph{e.g.}, aerosols) and trace gases with low ionization potential.

\begin{figure}[htp]
\begin{center}
\includegraphics[width=0.5\textwidth]{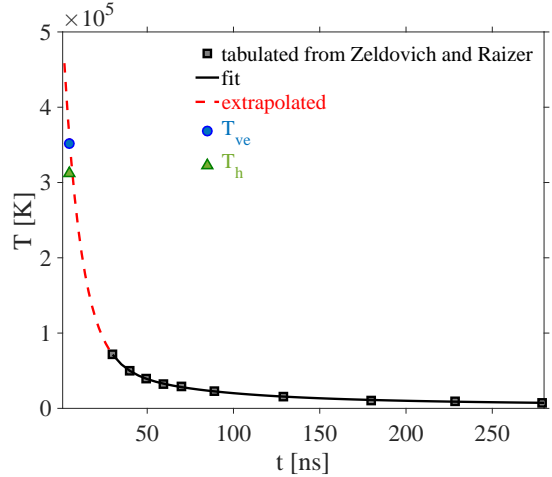}
\vspace*{-0.2cm}
\caption{\label{fig:intro_temp}Evolution of spark temperature during breakdown and post-breakdown stages: maximum free-electron vibronic ($T_\mathrm{ve}$) and heavy-particle ($T_\mathrm{h}$) temperatures predicted by the authors for case~2 (see Sec.~\ref{sec:dynamics}) at \SI{5}{\nano\second} into the discharge (circle and triangle, respectively), temperature behind the shock wave reported by Harilal \emph{et al.} \cite{harilal2015lifecycle} (square symbol, tabulated values from Zeldovich and Raizer), curve-fit of Harilal \emph{et al.}  in the post-discharge phase (solid black line), and extrapolation of the curve-fit into the discharge phase (red broken line).}
\end{center}
\end{figure}

The final outcome of laser-induced breakdown (LIB) is the formation of a transient high-temperature and strongly ionized plasma. For nano-second pulses considered here, maximum temperatures on the order of $10^5$ up to $10^6$ \SI{}{\kelvin} are to be expected during the discharge. This conclusion was first reached independently by both De Michelis \cite{DeMichelis_IEEE_1969}, who concentrated on LIB in atmospheric air (using a ruby laser $h_{\textsc{p}} \nu = \SI{1.78}{eV}$), and by Zel'dovich and Raizer \cite{Zeldovich_JETP_1965,Zeldovich_book_1967}. These estimates have been corroborated by both experiments and calculations. For instance, in the work by Bradley \emph{et al.} \cite{bradley2004fundamentals}, where a Q-switched Nd:YAG laser was used to promote breakdown in both air and isooctane-air mixtures, free-electron temperatures above $10^5$ \SI{}{\kelvin} were inferred based on the blast-wave theory. Other authors, such as Phuoc and White\cite{phuoc2002optical}, report temperatures larger than $10^5$ \si{kelvin}. The writers of this paper \cite{munafo_JCP,alberti_JPhysD,alberti_dual} performed physics based numerical simulations which led to peak temperatures around $\SI{5e5}{\kelvin}$. After the discharge, the plasma expands and propagates through the ambient gas. This is accompanied by a significant temperature decrease as a result of conservation of energy and growth of the front of the laser-induced blast wave. With time, the latter degenerates into an approximately spherical sound wave. A typical temperature vs time evolution plot for a nano-second laser discharge is shown in Fig. ~\ref{fig:intro_temp}. There, the values from Harilal \emph{et al.} \cite{harilal2015lifecycle} (who fitted spark temperature behind the shock wave data by Zel'dovich and Raizer; Harilal \emph{et al.} use a \SI{1064}{\nano\meter} pulse at \SI{55}{\milli\joule} from a Q-switched Nd:YAG laser Continuum Surelite II with a duration of \SI{6}{\nano\second} FWHM, and a focal spot size of \SI{10}{\micro\meter}) are extrapolated into discharge phase and compared with the maximum temperatures predicted by the authors at $t = \SI{5}{\nano\second}$ for LIB in air. The overall agreement is good and further confirms the above figures on temperature values to be expected during the discharge phase.      

Modeling of LIB in gases is a complex multi-physics problem, requiring a self-consistent coupled solution of the equations governing hydrodynamics and radiation. Under most situations, the gas onto which the laser is focused is collision dominated and, in light of this, may be treated using a fluid description (\emph{e.g.}, Navier--Stokes equations). Emission, absorption and scattering of light are taken into account via the Radiative Transfer Equation (RTE), which is the equivalent of Boltzmann's equation for a photon gas \cite{oxenious} (see \ref{appendixA}). The inclusion of finite-rate (\emph{i.e.}, non-equilibrium) collisional and radiative processes (\emph{e.g.}, ionization, absorption) is accomplished by means of either multi-temperature \cite{kandala2004numerical,tropina2017mathematical,Shiva_POP_2017,Mahamud_2018,munafo_JCP,alberti_JPhysD,alberti_dual} or State-to-State models \cite{colonna2001modelling,Casavola_ApplOpt_2003}. For CFD applications, the former is often the preferred choice due to large computational cost associated to tracking the population of hundreds to thousands atomic/molecular \emph{bound} states as required by a State-to-State approach. 

In their previous works, the authors developed and validated a predictive LIB model where the RTE, which accounted for attenuation and propagation of the laser beam, is coupled to the Navier--Stokes equations for a multi-component plasma \cite{munafo_JCP,alberti_JPhysD,alberti_dual,alberti_thesis}. In this computational framework non-equilibrium phenomena are taken into account via a two-temperature model (though State-to-State models may be also used in light of a general and flexible implementation). As opposed to what is usually done in the literature \cite{kandala2004numerical,tropina2017mathematical,Shiva_POP_2017,Mahamud_2018,dors2003computational,dumitrache2020gas}, both multiphoton ionization (MPI) and inverse Bremsstrahlung (IB) are self-consistently coupled with the hydrodynamics. This avoids using an initial artificial plasma and, more importantly, was crucial in reproducing the experimentally observed plasma kernel dynamics. In particular, during plasma formation, MPI is the guiding process of the plasma waves. This is opposed to the common view according to which MPI is considered important only at early times for the production of \emph{priming} electrons (\emph{e.g.}, around pico-second timescales), and becomes negligible thereafter. The LIB model developed by the authors has been extensively validated for single-mode pulses (\emph{e.g.}, Gaussian time evolution) for a broad range of beam energy and wavelength, and ambient conditions. Overall good agreement with experiments has been reported for both the discharge \cite{alberti_JPhysD} and the post-discharge phases \cite{alberti_dual}.

During the discharge, spatially periodic plasma kernel structures are observed for both femto-second and nano-second pulses in both single- and multi-mode configurations \cite{nishihara2018influence, sprangle1990nonlinear, mora1997kinetic, cooley2004modeling}. Moreover, for multi-mode lasers, an enhancement of electrons produced via MPI has also been predicted \cite{carusotto1973incoherent} and observed \cite{lecompte1974experimental}. Several mechanisms have been proposed for describing the onset and dynamics of the periodic structures such as self-focusing \cite{alcock1970breakdown}, lens aberration \cite{evans1969lens} and hydrodynamic instabilities \cite{margolin1999generation}.  For intense, short pulses (\emph{e.g.}, $10^{13} \, \si{\watt/\cm^{2}}$) self-focusing has been confirmed as the primary mechanism \cite{kaw1969nonlinear}, but these structures are also observed at lower intensities of nano-second discharges. Nishihara \emph{et al.} \cite{nishihara2018influence} reported that the spatial wavelength between the periodic structures seems to agree with hydrodynamic instabilities generated by the interaction at the plasma surface between excited and transmitted plasma \emph{waves}. It has been suggested that lateral radiative heat flux in the breakdown region and the rapid plasma expansion combined together might result in in-homogeneous refraction, which can then produce periodic structures \cite{margolin1999generation}. The period in this case would depend on the laser wavelength and the angle of inclination of the Bessel beam \cite{margolin1999generation}. However, Nishihara \emph{et al.} \cite{nishihara2018influence}  observed that the periodic structure of the electron density distribution was related to the laser mode-beating frequency, suggesting that the multi-mode pulse either seeds instabilities or directly generates periodic kernel structures that match the beating mode.


The purpose of this work is to investigate the generation mechanism of plasma periodic structures observed in multi-mode nano-second pulses \cite{nishihara2018influence} by means of the self-consistent LIB model developed by the authors \cite{munafo_JCP,alberti_JPhysD,alberti_dual}.    

The paper is structured as follows. Section \ref{sec:model} describes the physical model and the numerical method. In Section \ref{sec:res}, the results for multi-mode LIB in air are first compared against experiments for validation. This is followed by an in-depth analysis of the generation mechanism of the plasma periodic structures. Conclusions are summarized in  Sec.~\ref{sec:conc}.
\section{Physical model and numerical method}  \label{sec:model}
The gas onto which the laser is focused is air made of \SI{79}{\percent} $\mathrm{N_2}$ and \SI{21}{\percent} $\mathrm{O_2}$ at ambient conditions. The laser-induced plasma is modeled as mixture of ideal gases (\emph{e.g.}, no pressure ionization \cite{Zeldovich_book_1967}) and accounts for chemical components resulting from dissociation and ionization of molecular nitrogen and oxygen:
$$
\mathcal{S} = \{\mathrm{e}^-,\, \mathrm{N},\, \mathrm{O},\, \mathrm{N}_2,\, \mathrm{O}_2,\, \mathrm{NO},\, \mathrm{N}^+,\, \mathrm{N}^{2+},\, \mathrm{N}^{3+},\, \mathrm{N}^{4+},
$$
\vspace{-0.5cm}
$$
\mathrm{N}^{5+},\, \mathrm{O}^+,\, \mathrm{O}^{2+},\, \mathrm{O}^{3+},\, \mathrm{O}^{4+},\, \mathrm{O}^{5+},\, \mathrm{N}_2^+,\, \mathrm{O}_2^+,\, \mathrm{NO}^+\} \, .
$$
Here only positively charged ions are considered.  
\paragraph{Hydrodynamics} The non-equilibrium hydrodynamics are described based on a two-temperature model, where the heavy-particle temperature, $T_{\mathrm{h}}$, is associated with the roto-translational degrees of freedom of heavy-particles (\emph{i.e.}, atoms and molecules), whereas the free-electron-vibronic temperature, $T_{\mathrm{ve}}$, is associated with translation of free-electrons and vibration of molecules and electronic excitation of heavy-particles. The flow governing equations are \cite{munafo_JCP,Gnoffo}:

\begin{IEEEeqnarray}{rCl}
\frac{\partial \rho_s}{\partial t} + \nabla \cdot \left( \rho_s \mathbf{u} \right) & = & -\nabla \cdot \mathbf{J}_s + \omega_s, \quad s \in  \mathcal{S}, \label{gas:mass_conservation_heavy} \\
\frac{\partial \rho \mathbf{u}}{\partial t} + \nabla \cdot \left( \rho \mathbf{u} \mathbf{u} + p \, \mathsf{I} \right) &= & \nabla \cdot \stau, \label{gas:momentum_conservation}  \\
\frac{\partial \rho E}{\partial t} + \nabla \cdot (\rho H  \mathbf{u} ) & = & \nabla \cdot ( \stau \mathbf{u} - \mathbf{q}) +  \Omega^{\textsc{r}}, \label{gas:energy_conservation} \\
\frac{\partial \rho e_{\text{ve}}}{\partial t} +\nabla \cdot \left(\rho e_{\text{ve}}  \mathbf{u}\right) & = & - \nabla \cdot  \mathbf{q}_{\text{ve}} - p_{\mathrm{e}}  \nabla  \cdot \mathbf{u} + 
\label{gas:vibr_energy_conservation} 
\end{IEEEeqnarray}
\vspace{-0.70cm}
\begin{eqnarray*}
\qquad \qquad \qquad + \Omega_{\mathrm{ve}}^{\textsc{c}} + \Omega_{\text{ve}}^{\textsc{r}},
\end{eqnarray*}

where the `ve' subscript identifies quantities related to the free-electron-vibronic degrees of freedom alone. The symbols in Eqs.~\eqnref{gas:mass_conservation_heavy}-\eqnref{gas:vibr_energy_conservation} have their usual meaning \cite{munafo_JCP,alberti_JPhysD,Gnoffo,alberti_dual}. The $\rho_s$ are the partial densities, from which the gas density may obtained via $\rho = \sum_{s \in \mathcal{S}} \rho_s$. The $p$ and $p_{\mathrm{e}}$ symbols denote, respectively the gas and free-electron pressure. The capital letter $E$ stands for the total energy per unit-mass. The former accounts for thermal, formation and kinetic contributions and may be used to compute the total enthalpy based on its definition, $H = E + p/\rho$. The greek symbols $\omega$ and $\Omega$ represent, respectively, mass and energy production terms due to collisional (C) and radiative (R) processes. The boldface quantities $\mathbf{u}$, $\mathbf{J}_s$ and $\mathbf{q}$ denote the mass-averaged velocity, the mass diffusion and the heat flux vectors, respectively, whereas the symbols $\mathsf{I}$ and $\stau$ stand for the identity and viscous stress tensors, respectively. The collisional kinetic processes included in the model are: dissociation by heavy-particle and electron impact, ionization by electron impact, particle exchange (e.g. Zel’dovich reactions), charge exchange, and associative ionization reactions. Details on the constitutive relations for thermodynamics, transport and kinetics for the governing equations \eqnref{gas:mass_conservation_heavy}-\eqnref{gas:vibr_energy_conservation} are reported elsewhere \cite{munafo_JCP,alberti_thesis}.
\paragraph{Radiation} The evolution of the monochromatic intensity of the collimated beam, $I_{\lambda}$, where $\lambda$ stands for its wavelength, is governed by the Radiative Transfer Equation (RTE) \cite{oxenious}. Since the collimated radiation (i.e., the laser beam) is expected to represent the prominent role in the radiative energy transfer for the laser under investigation, and since the time scales of laser pulse and hydrodynamics are much slower of light propagation, we can assume steady-state (\emph{i.e.}, $\pa ()/\pa t = 0$) and neglect emission, scattering and refraction. Under those assumptions, the RTE reduces to Beer--Lambert's law:
\begin{equation}
\mathbf{n} \cdot \nabla I_{\lambda} = - \kappa_{\lambda}  I_{\lambda},
\label{rad:rte} 
\end{equation}
where $\kappa_{\lambda}$ is the monochromatic absorption coefficient corrected for stimulated emission, and $\mathbf{n}$ is the direction cosine along the line of sight. 

The radiative processes taken into account are: multiphoton ionization (MPI) and inverse Bremsstrahlung (IB). MPI rate coefficients  are evaluated based on Grey Morgan's formula \cite{morgan1975laser}, whereas the IB absorption coefficient is computed using Kramers' formula for electron--ion interactions \cite{Zeldovich_book_1967} and the model by Kroll and Watson \cite{kroll1972theoretical} for electron--neutral interactions. More details are reported elsewhere \cite{munafo_JCP,alberti_JPhysD}.
\paragraph{Numerical Method.} The axially symmetric form of Eqs.~\eqnref{gas:mass_conservation_heavy}-\eqnref{gas:vibr_energy_conservation} is solved using the \textsc{hegel} plasma solver \cite{munafo_JCP}. The flow governing equations are first discretized in space based on the cell-centered Finite Volume method. Inviscid fluxes are evaluated using van Leer's flux vector splitting in conjunction with a second-order upwind-biased \textsc{MUSCL} reconstruction. Diffusive fluxes are computed using Green-Gauss's formula to evaluate face-averaged gradients. The spatial dicretization is followed by a temporal integration accomplished via an Implicit-Explicit (IMEX) dual-time-stepping method. The RTE \eqnref{rad:rte} is solved based on a flux-tube formulation \cite{munafo_JCP}.
\section{Results}  \label{sec:res}
The simulation set-up (\emph{e.g.}, grid, laser beam and focusing system parameters) for the various cases are introduced in Sec.~\ref{sec:set-up}. This is followed by validation via comparison against experiments in Sec.~\ref{sec:Validation}. Finally, the generation mechanism of plasma periodic structures is investigated in Sec. \ref{sec:dynamics}.
\subsection{Simulation set-up} \label{sec:set-up}
\begin{figure}[htp]
\includegraphics[width=0.5\textwidth]{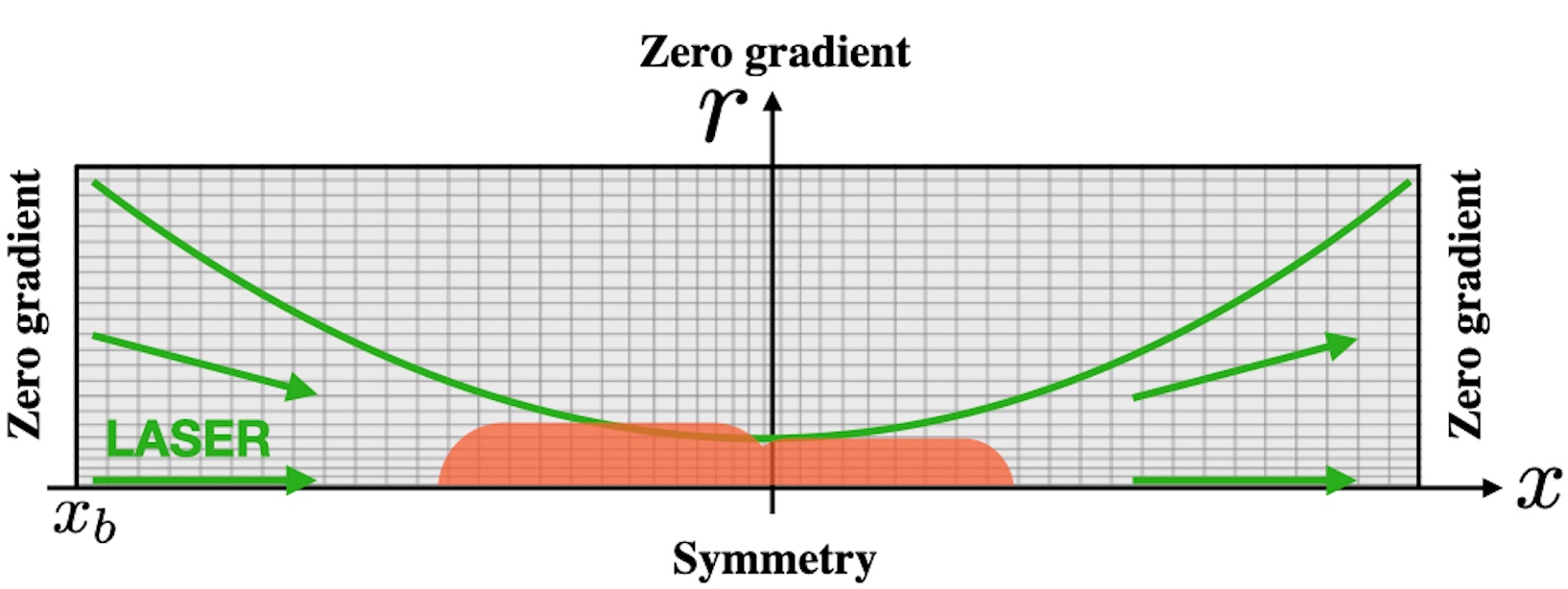}
\vspace*{-.2cm}
\caption{Schematics of the simulation set-up.}
\label{fig:schematics_setup}
\end{figure}

A rectangular (axisymmetric) computational domain is discretized using a two-dimensional Cartesian structured grid along the $x$ and $r$ coordinates (see Fig.~\ref{fig:schematics_setup}). The focal plane is taken at the origin. The beam propagates along the positive $x$ direction. 

The boundary conditions for the hydrodynamic equations \eqnref{gas:mass_conservation_heavy}-\eqnref{gas:vibr_energy_conservation} are: symmetry on the optical (\emph{i.e.}, $x$) axis, and zero-gradient at the other boundaries. For the laser beam, the intensity at the incoming boundary $I_{\lambda}(x_b,r,t)$, placed at $x=x_b\, < 0$, is prescribed as:
\begin{equation}\label{eq:lasr_BC2}
I_{\lambda} = I_0  \, \exp \left( -  \frac{r^2}{2 \sigma_r^2} \right) g_{mb}(t) \exp \left[-\frac{(t-t_p/2)^2}{2 \sigma_t^2} \right],
\end{equation}
where the mode-beating function $g_{mb}(t) = \left[ 1 \, + A_{mb}  \,  \sin (2 \pi f_{mb} t) \right]$, with $A_{mb}$ and $f_{mb}$ being, respectively, the corresponding non-dimensional amplitude and frequency, is introduced to modulate the Gaussian time dependence. The temporal standard deviation is obtained from $\sigma_t = \mathrm{FWHM}/2\sqrt{2\ln 2}$, where the Full-Width at Half Maximum (FWHM) is either taken from experiments or imposed. The radial standard deviation $\sigma_r$ is determined based on the focal radius, beam diameter at focusing lens and focal length, assuming a diffraction limited par-axial beam. The procedure is discussed in Ref \cite{munafo_JCP}. The duration of the pulse is taken as $t_p = 6 \, \sigma_t$. The $I_0$ constant is determined by imposing that the energy entering the domain at $x = x_b$ is equal to that delivered by the laser ($\varepsilon$):   
\begin{widetext}
\begin{equation}\label{eq:lasr_norm_I0}
\varepsilon = 2 \pi \!\! \int\limits_{0}^{+\infty} \int\limits_{ 0}^{t_p} I_{\lambda}(x_b,r,t) \, r \, dr \, dt = 2\pi \sigma^2_r  I_0  \int\limits_{ 0}^{6 \sigma_t} g_{mb}(t) \exp \left[-\frac{(t-3\sigma_t)^2}{2 \sigma_t^2} \right] dt.
\end{equation}	
\end{widetext}

The flowfield is initialized in a quiescent and uniform state with ambient pressure and temperature equal to $p_a = \SI{101325}{\pascal}$ and $T_a = \SI{288}{\kelvin}$, respectively, to match the conditions of Ref. \cite{nishihara2018influence}. In all simulations the beam wavelength is $\lambda = \SI{532}{\nano\meter}$.
\subsection{Model validation}\label{sec:Validation}
\begin{figure}[htp]
\begin{center}
\includegraphics[width=0.5\textwidth]{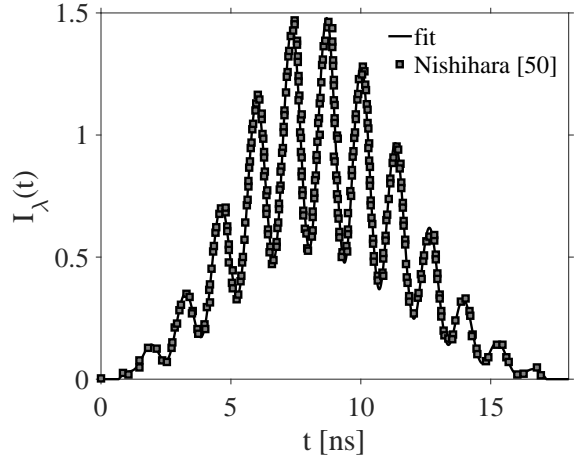}
\caption{Multi-mode waveform: comparison between fitting formula (\emph{i.e.}, bracketed term in Eq. (\ref{eq:temporal_fit})) and calculated experimental waveform of Nishihara \emph{et al.} \cite{nishihara2018influence}.}
\label{fig:multi_fitted}
\end{center}
\end{figure}

The non-equilibrium LIB model developed by the authors has been extensively validated through comparison against experiments for single-mode pulses \cite{alberti_JPhysD,alberti_dual}. Here validation is pursued further considering a multi-mode pulse. Experimental data are taken from the work by Nishihara \emph{et al.} \cite{nishihara2018influence} who investigated LIB in atmospheric air at three different energies: \SI{30}{\milli\joule}, \SI{60}{\milli\joule} and \SI{100}{\milli\joule}. For the sake of a more consistent comparison, the temporal part of the incoming intensity (\ref{eq:lasr_BC2}) is replaced here by a multi-mode waveform obtained by fitting the data from Nishihara \emph{et al.} (\emph{i.e.}, from Fig. 3(b) of \cite{nishihara2018influence}): 
\begin{widetext}
\begin{equation}\label{eq:temporal_fit}
I_{\lambda}(x_b,r,t) = I_0  I_{\lambda}(x_b,r)  I_{\lambda}(t)  = I_0   \exp \left( -  \frac{r^2}{2 \sigma_r^2} \right) \max \left[ \sum_{k=1}^8 a_k  \sin (b_k  t  +  c_k),   0 \right],
\end{equation}
\end{widetext}
where the $a_k$, $b_k$ and $c_k$ fit coefficients (for $\num{0} \leq t \leq \SI{18}{\nano\second}$) are given in Table~\ref{tab:fit}. The $I_0$ constant is always determined based on Eq. (\ref{eq:lasr_norm_I0}). Figure \ref{fig:multi_fitted} compares the fitted multi-mode waveform (\emph{i.e.}, the bracketed term in Eq. (\ref{eq:temporal_fit})) with the experimental profile.

\begin{table}[!ht]
\begin{center}
\begin{tabular}{ c | c | c | c  }
$k $  &  $a_k$ &  $b_k \; [\SI{}{\giga\hertz}]$   & $c_k$  \\
\hline
\hline
1 & \num{0.8458} & \num{0.2029} & \num{-0.141}        \\
2 & \num{0.01616} & \num{1.121} & \num{19.08}  \\
3 & \num{0.008364} & \num{3.569} & \num{-12.63}   \\
4 & \num{0.1703} & \num{0.6289} & \num{-3.567}    \\
5 & \num{-0.06075} & \num{4.081}  & \num{-6.067}  \\
6 & \num{0.219} & \num{4.493}  & \num{-5.947}  \\
7 & \num{0.03229} & \num{5.355} & \num{-7.695}   \\
8 & \num{0.1842} & \num{4.9} & \num{-9.502}   \\
\hline
\end{tabular}
\vspace{0.25cm}
\caption{Fit coefficients for the multi-mode waveform.}\label{tab:fit}
\end{center}
\end{table}

In all simulations conducted for validation purposes, the beam diameter at the focusing lens (with focal length $f = \SI{100}{\milli\meter}$) is \SI{9.8}{\milli\meter}, whereas its focal radius is $w_f = \SI{7.3}{\micro\meter}$ (as computed in Ref. \cite{alberti_JPhysD}). The value of the focal radius is within the prediction by Nishihara \emph{et al.} \cite{nishihara2018influence}, who calculated a focal radius of $\SI{10}{\micro\meter}$ in the absence of fluid expansion using the razor edge method, and considered this result an over-estimate of the actual value \cite{nishihara2018influence,skinner1972measurement}.

\begin{figure}[htp]
\begin{center}
\includegraphics[width=0.45\textwidth]{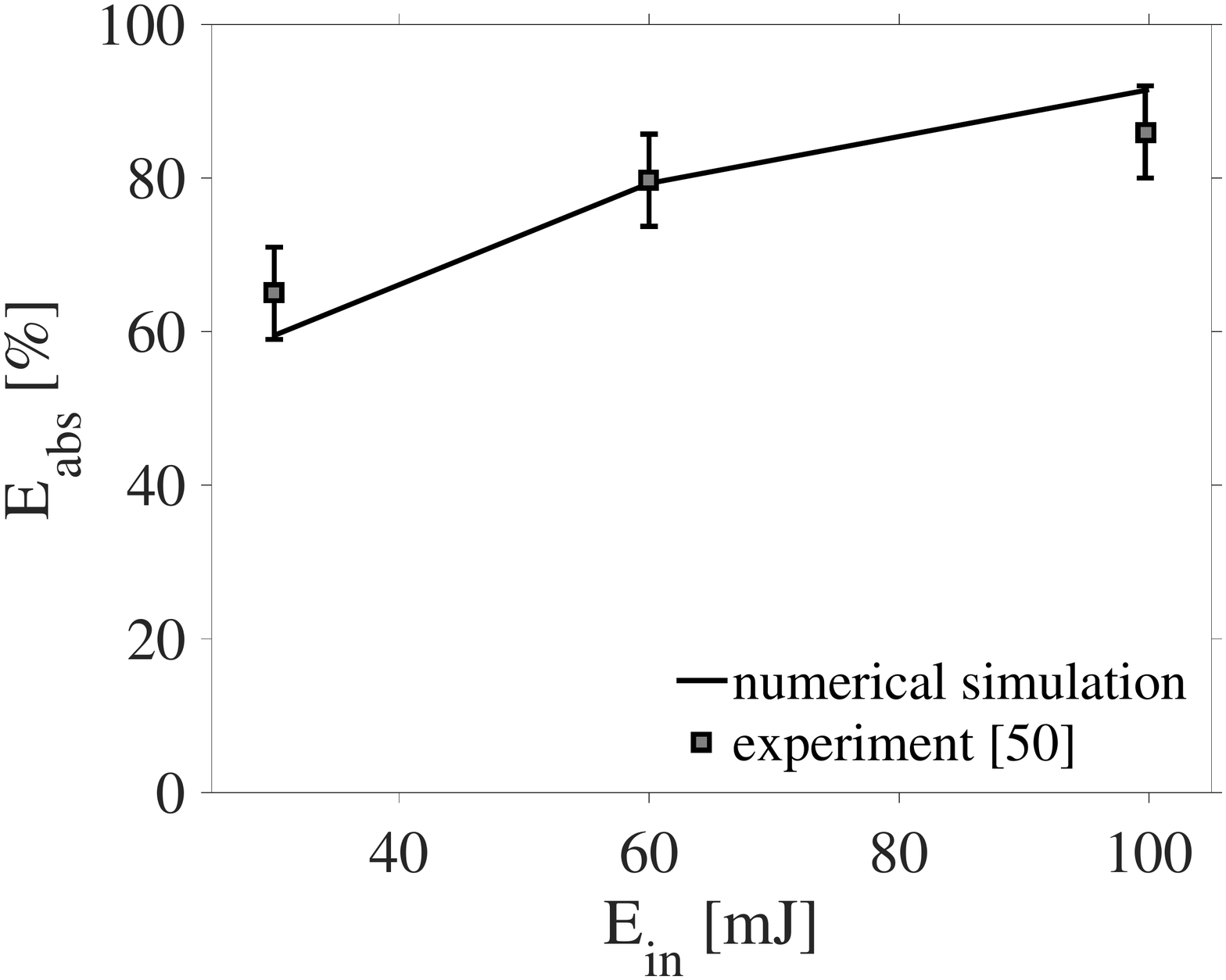}
\caption{Comparison between predicted and experimentally measured absorbed energy ($w_f~=~\SI{7.3}{\micro\meter}$, $f~=~\SI{100}{\milli\meter}$).}
\label{fig:multi_Eabs}
\end{center}
\end{figure}

\begin{figure}[htp]
\begin{center}
\includegraphics[width=0.45\textwidth]{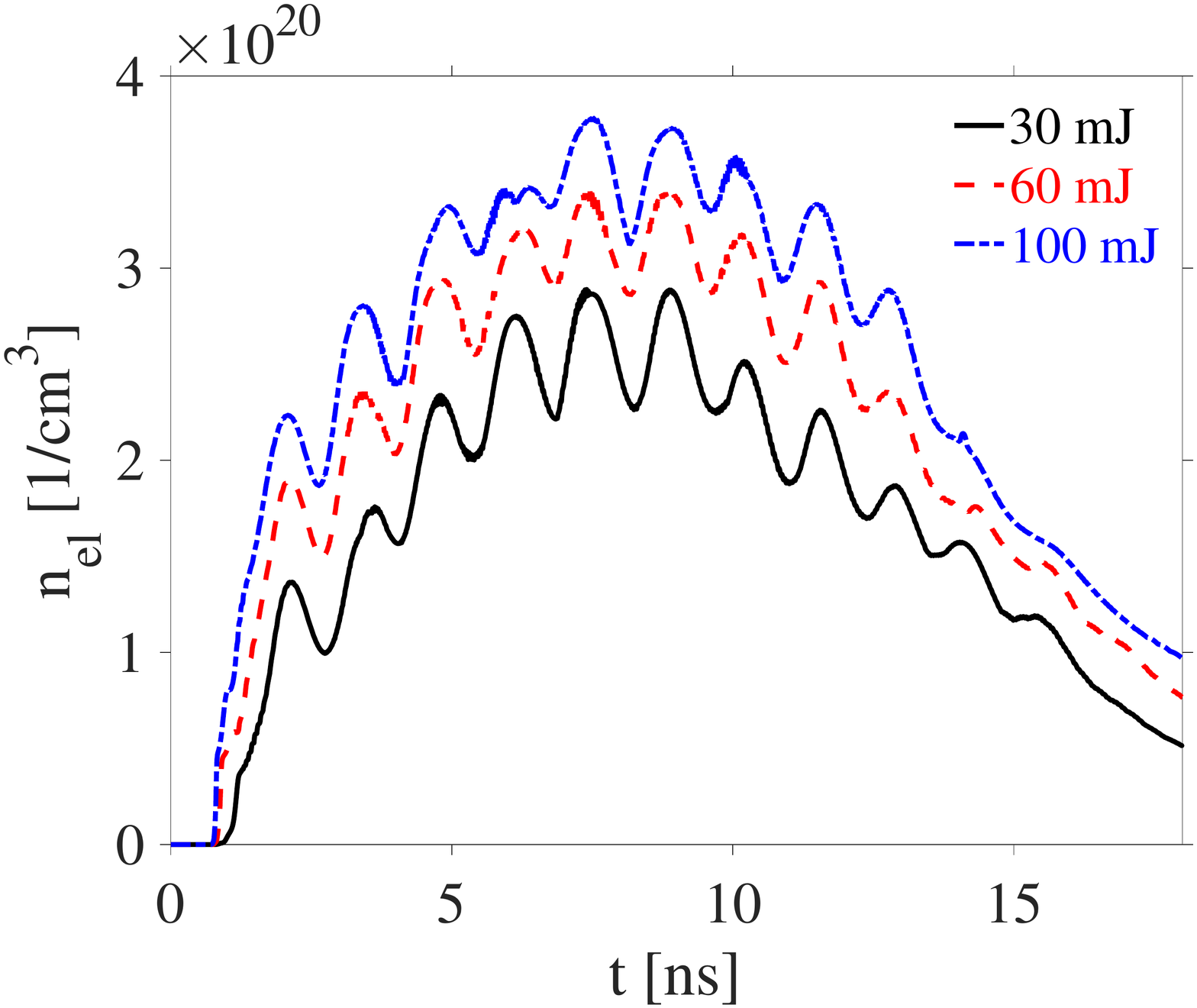}
\caption{Time-history of the maximum value of free-electron number density ($w_f~=~\SI{7.3}{\micro\meter}$, $f~=~\SI{100}{\milli\meter}$).}
\label{fig:multi_max_val_fitted}
\end{center}
\end{figure}

Figure \ref{fig:multi_Eabs} compares the computed absorbed energy with the measured non-transmitted energy. The agreement is good as the calculated absorbed energies fall within the experimental error bounds. At the end of the pulse ($t = \SI{18}{\nano\second}$), the maximum electron number density is between \SI{5.1e19}{} and \SI{9.6e19}{cm^{-3}} (depending on the laser energy, see Fig.~\ref{fig:multi_max_val_fitted}), consistently with what reported in Ref. \cite{nishihara2018influence}.

\begin{figure*}
\includegraphics[width=1.0\textwidth]{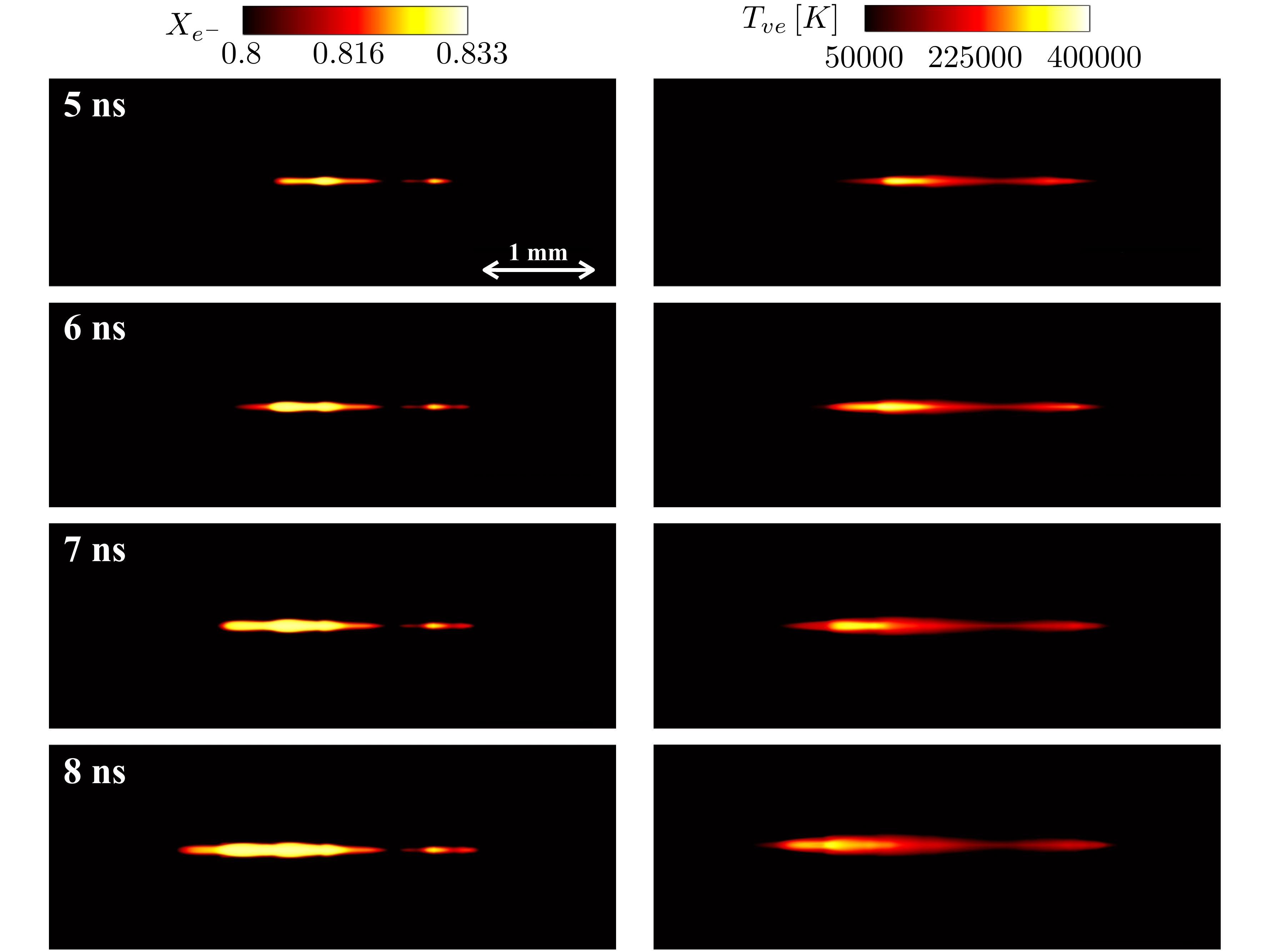}
\caption{Time-history of free-electron mole fraction (left) and temperature (right) for the $\SI{100}{\milli\joule}$ case ($w_f~=~\SI{7.3}{\micro\meter}$, $f~=~\SI{100}{\milli\meter}$).}
\label{fig:multi_contour_fitted}
\end{figure*}

\begin{figure*}[htp]
\hspace{-0.5cm}
\includegraphics[width=0.35\textwidth]{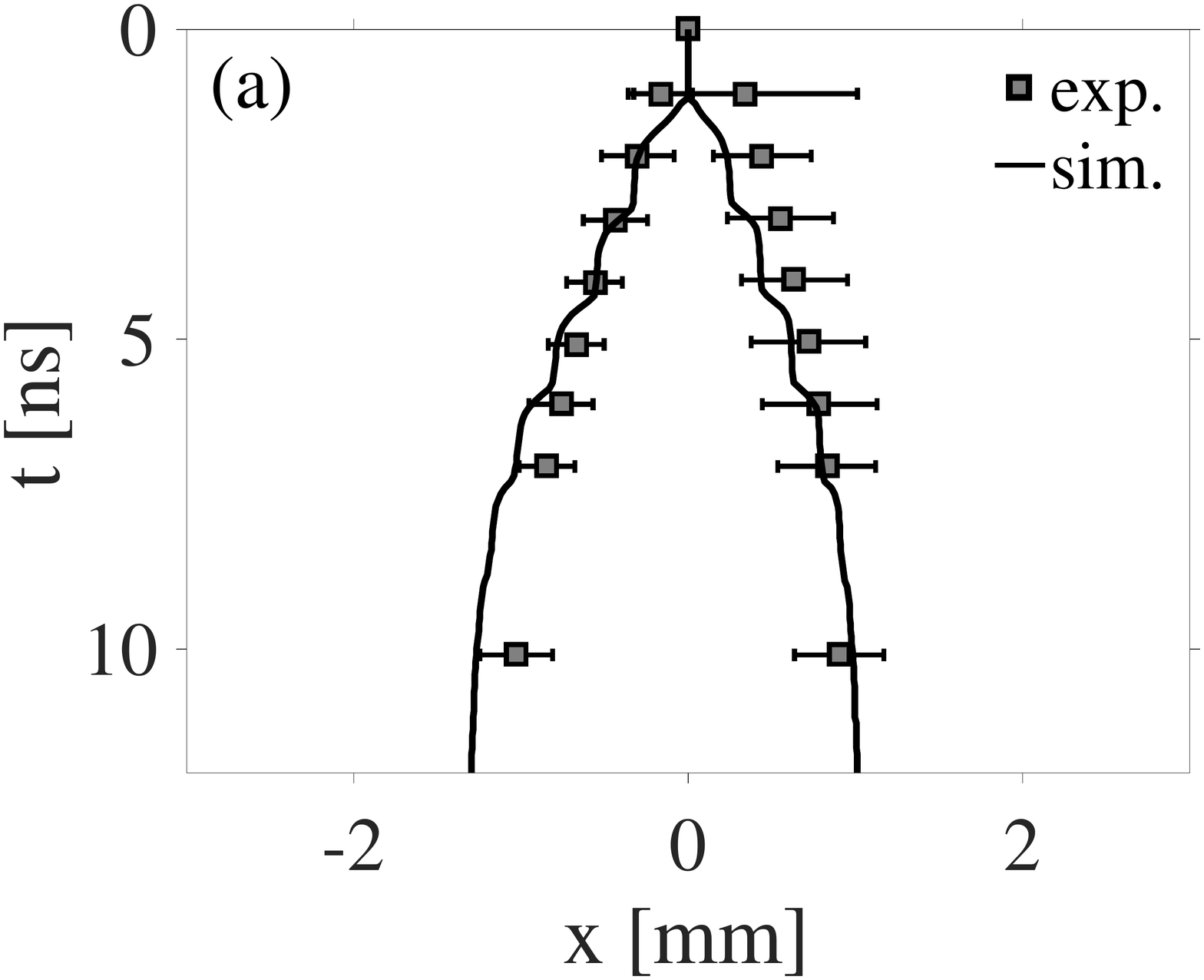}
\hspace{-0.5cm}
\includegraphics[width=0.35\textwidth]{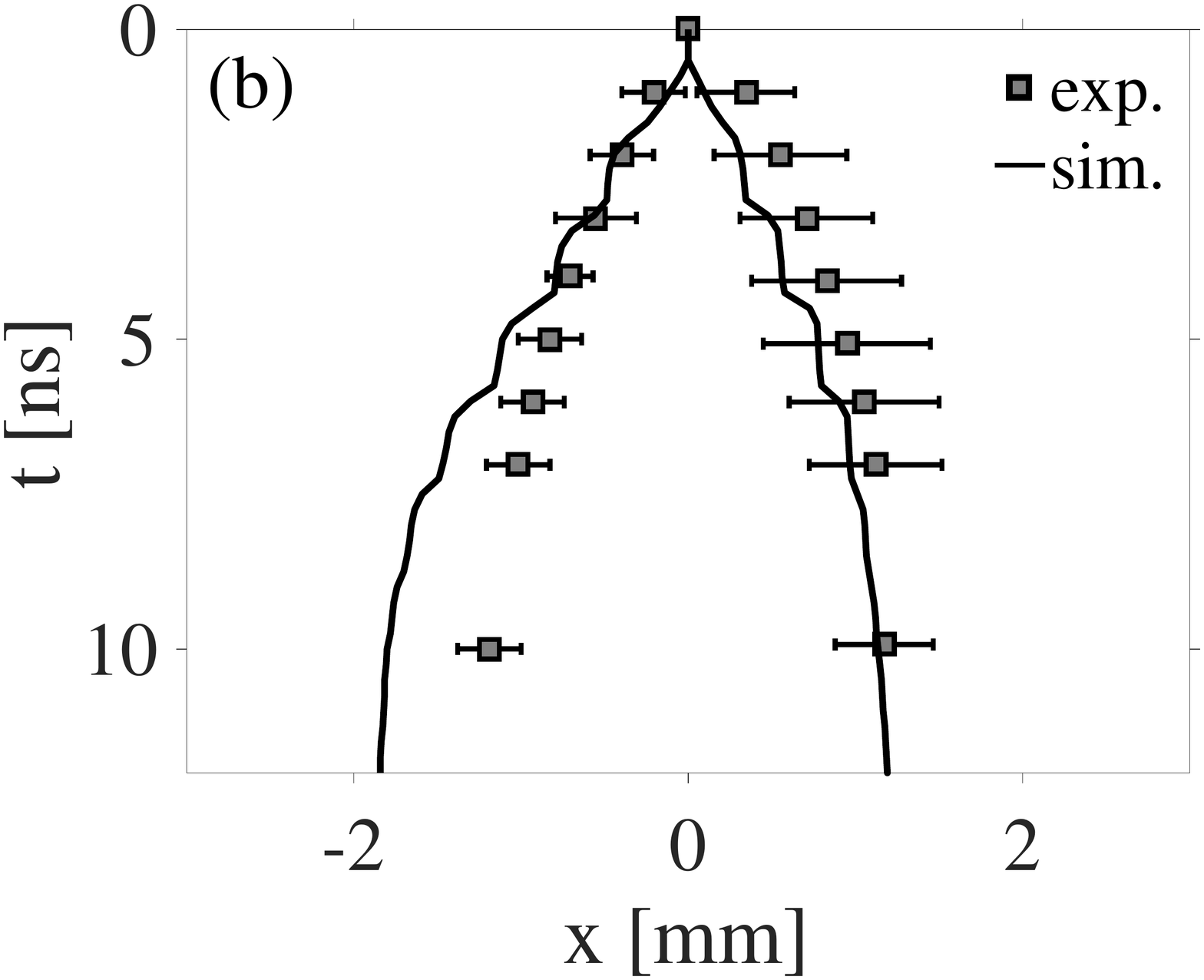}
\hspace{-0.5cm}
\includegraphics[width=0.35\textwidth]{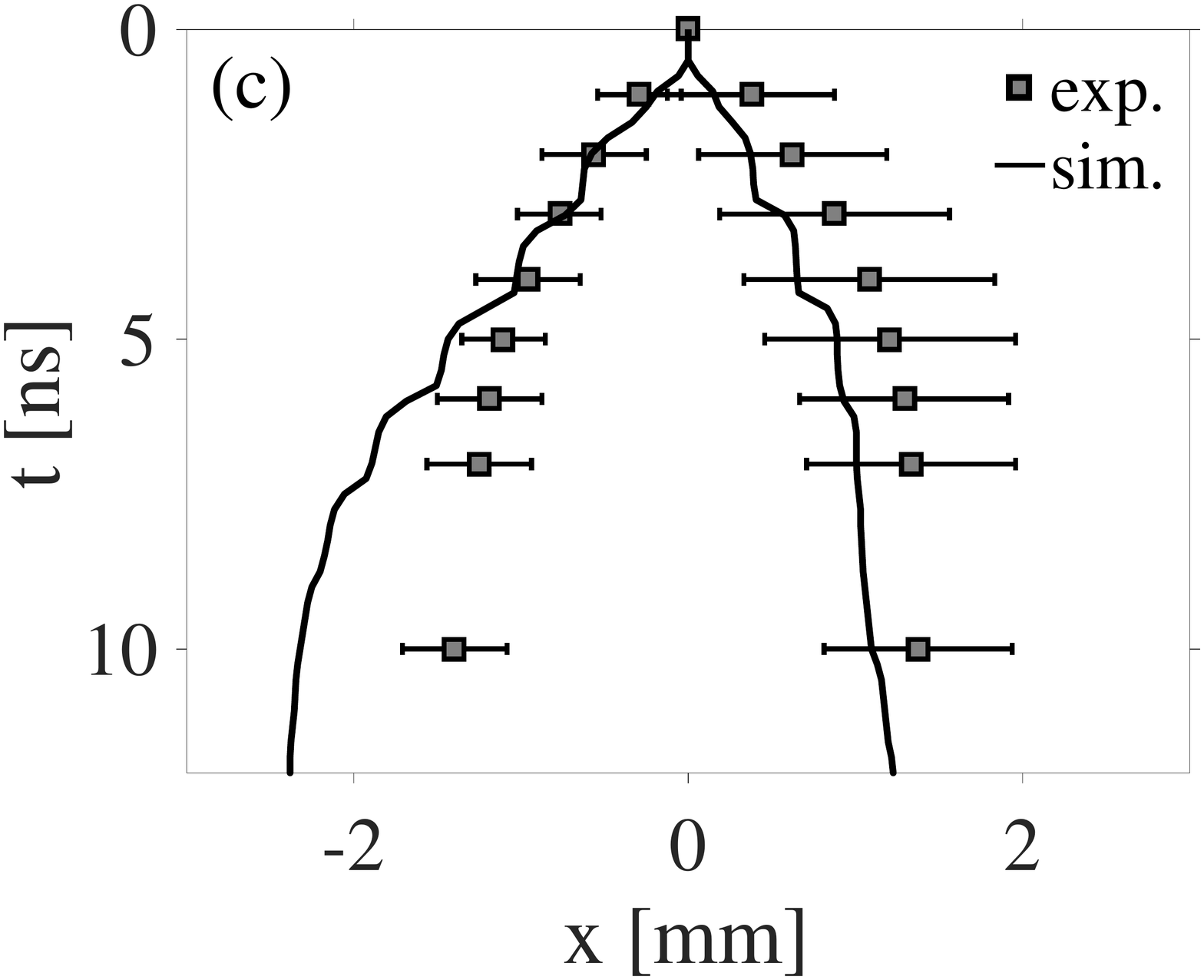}
\caption{$x$-$t$ diagram comparing the predicted and the measured temporal evolution of the plasma boundaries along the optical axis at different energies: (a) $\SI{30}{\milli\joule}$, (b) $\SI{60}{\milli\joule}$ and (c) $\SI{100}{\milli\joule}$ ($w_f~=~\SI{7.3}{\micro\meter}$, $f~=~\SI{100}{\milli\meter}$). Time increases top-to-bottom. Focal point at $x~=~\SI{0}{\milli\meter}$.}
\label{fig:multi_xt_diagram}
\end{figure*}

\begin{figure}[htp]
\begin{center}
\includegraphics[width=0.5\textwidth]{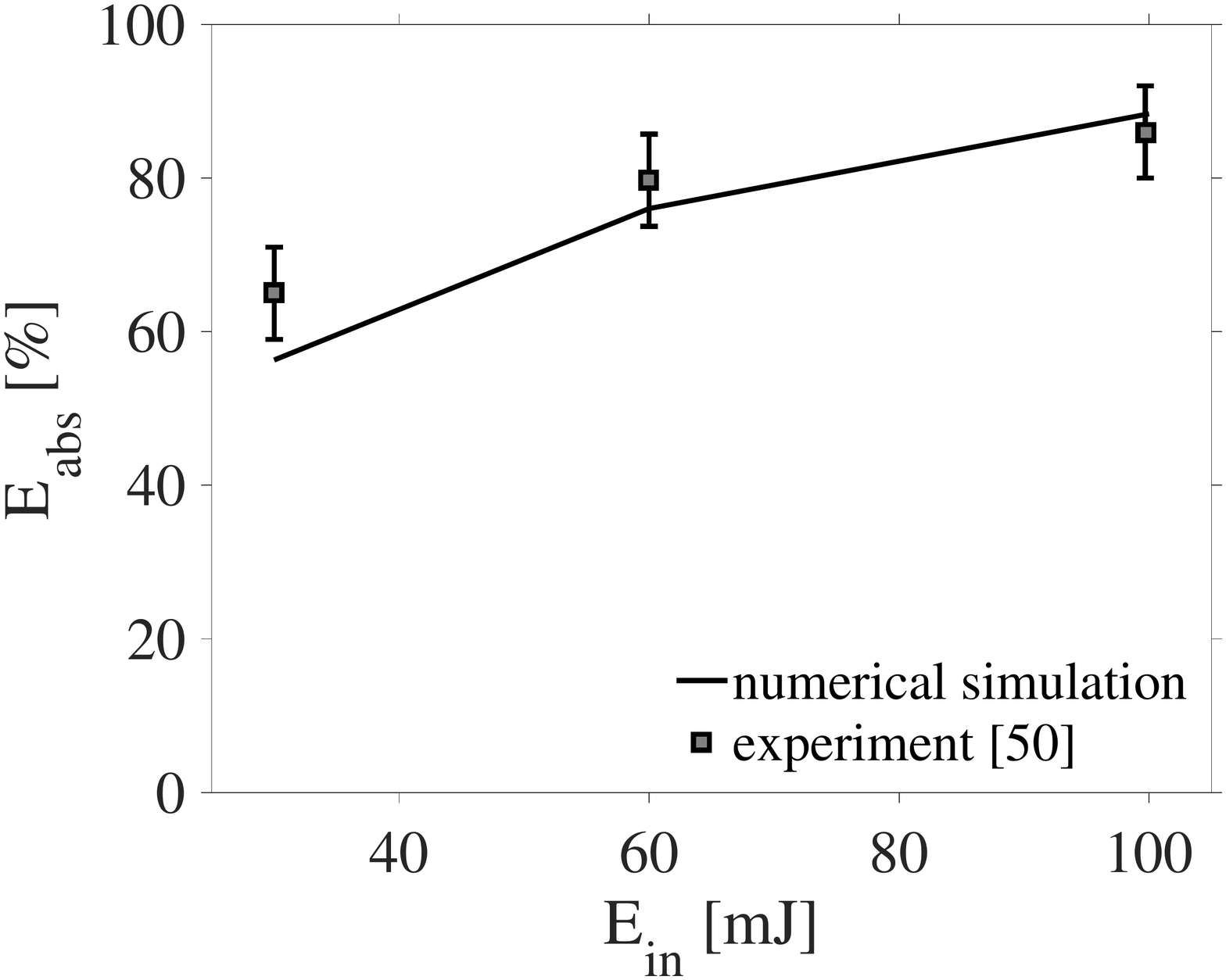}
\caption{Comparison between predicted and experimental absorbed energy ($w_f~=~\SI{10}{\micro\meter}$, $f ~=~\SI{80}{\milli\meter}$).}
\label{fig:multi_Eabs_fw}
\end{center}
\end{figure}

\begin{figure*}[htp]
\hspace{-0.5cm}
\includegraphics[width=0.35\textwidth]{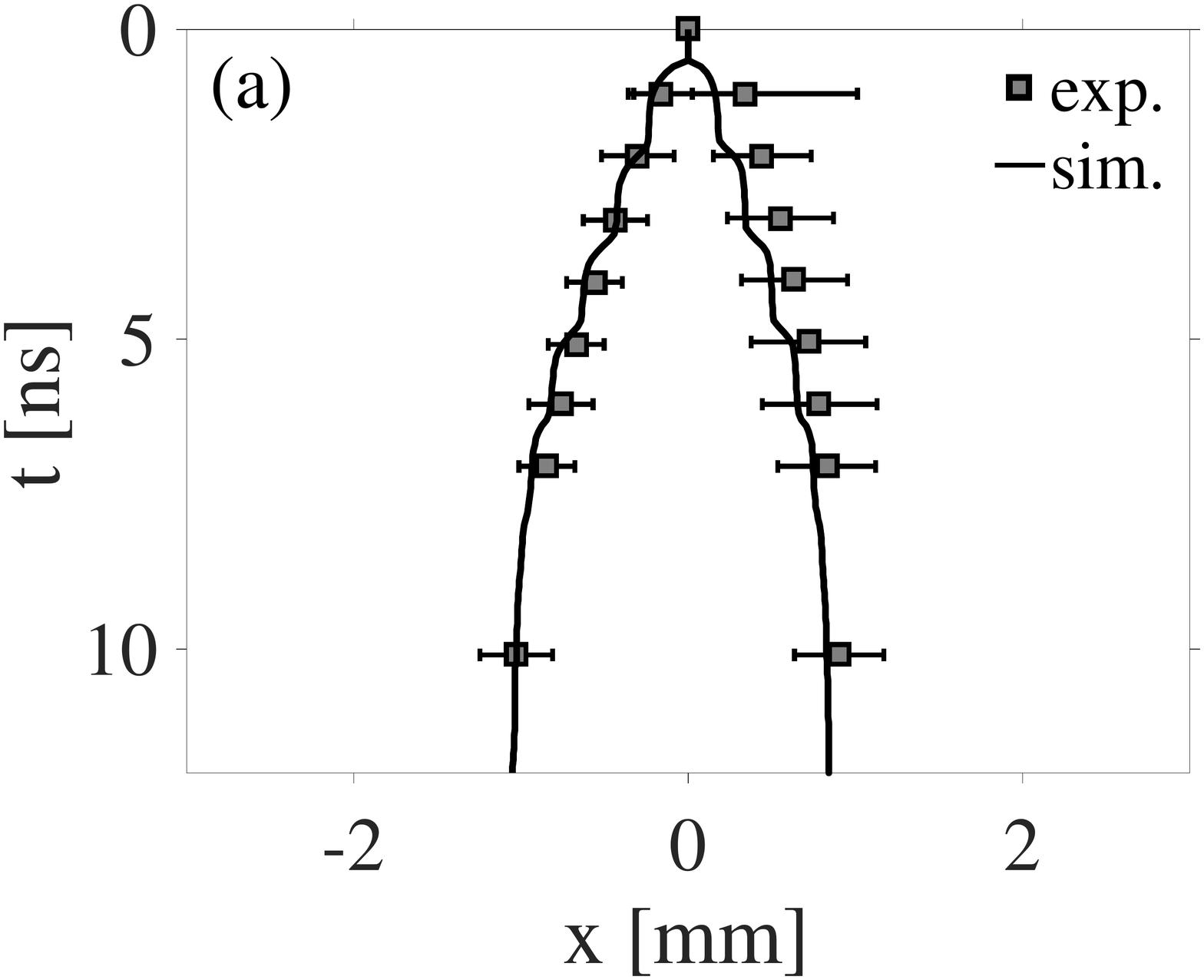}
\hspace{-0.5cm}
\includegraphics[width=0.35\textwidth]{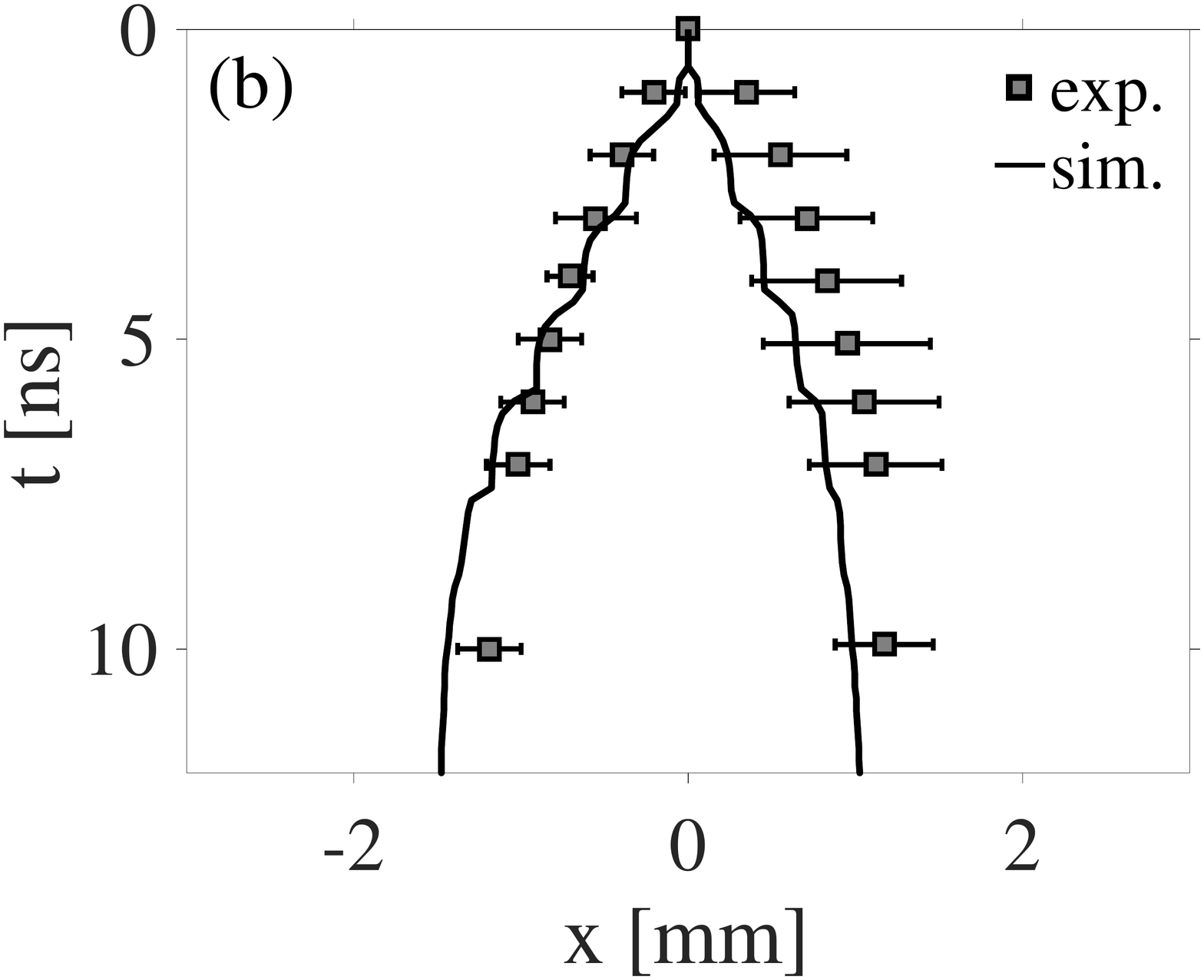}
\hspace{-0.5cm}
\includegraphics[width=0.35\textwidth]{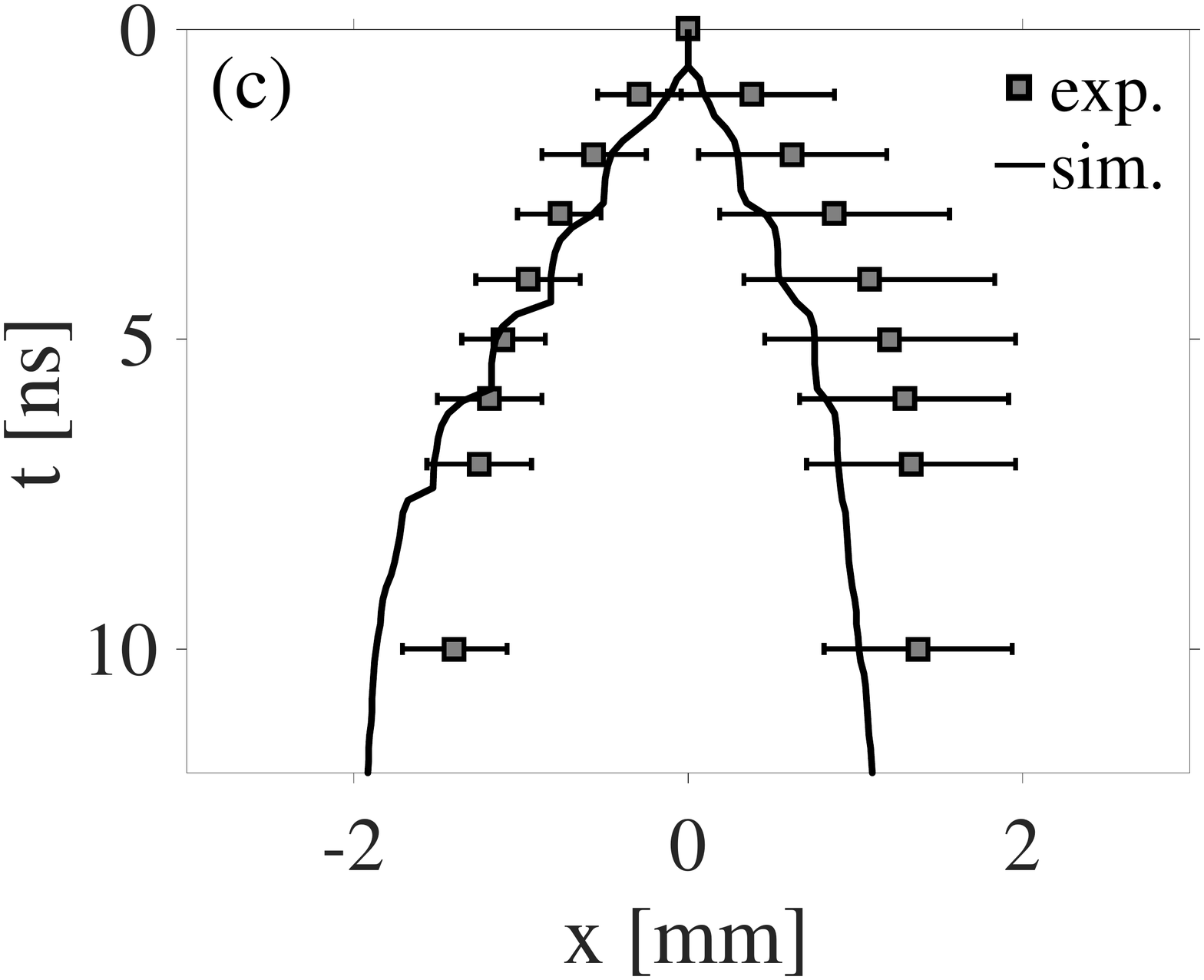}
\caption{$x$-$t$ diagram comparing the predicted and the measured temporal evolution of the plasma boundaries along the optical axis at different energies: (a) $\SI{30}{\milli\joule}$, (b) $\SI{60}{\milli\joule}$ and (c) $\SI{100}{\milli\joule}$ ($w_f~=~\SI{10}{\micro\meter}$, $f~=~\SI{80}{\milli\meter}$). Convention for the axes is the same as of Fig. \ref{fig:multi_xt_diagram}.}
\label{fig:multi_xt_diagram_fw}
\end{figure*}

\begin{figure*}[htp]
\begin{center}
\includegraphics[width=0.875\textwidth]{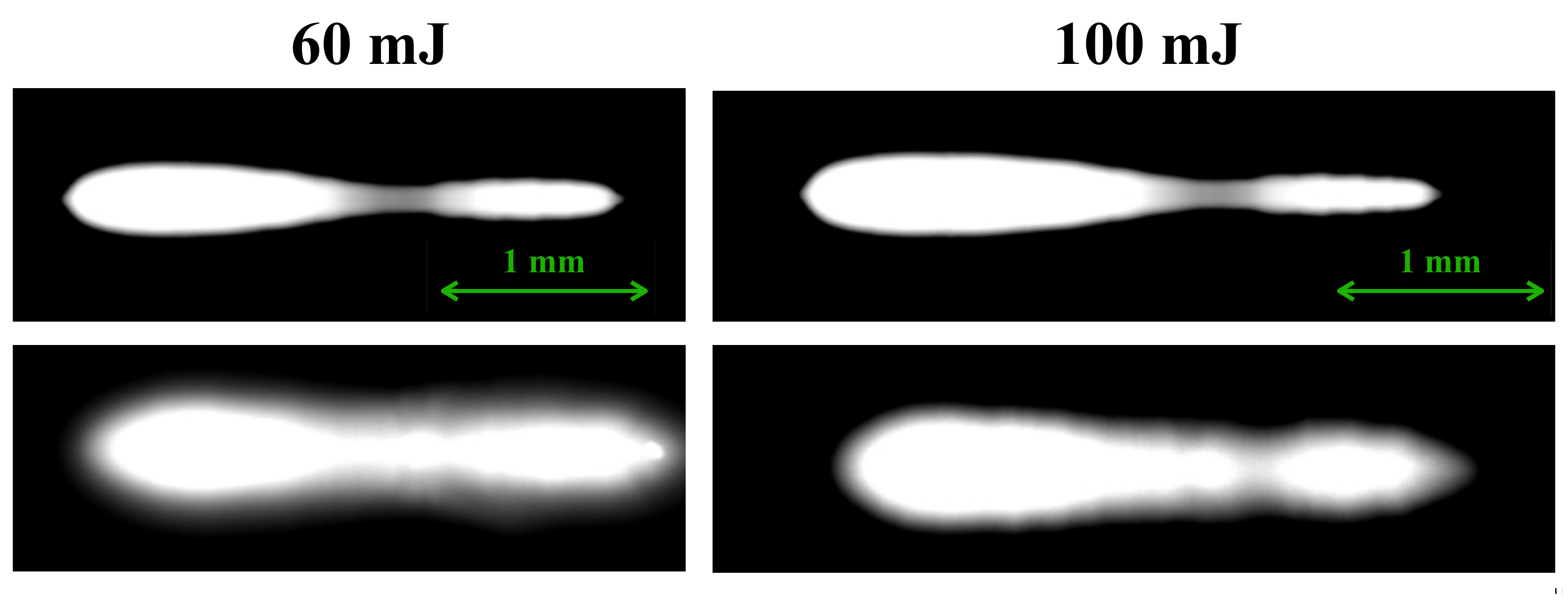}
\caption{Plasma emission images at the end of the pulse ($t~=~\SI{18}{\nano\second}$): top simulations  ($w_f~=~\SI{10}{\micro\meter}$, $f~=~\SI{80}{\milli\meter}$), bottom experiments.}
\label{fig:multi_emission_fw}
\end{center}
\end{figure*}

For a single-mode pulse the plasma kernel forms at the point of maximum intensity (\emph{i.e.}, at the focal point). The  \emph{priming} electrons formed there trigger the cascade ionization, and the plasma becomes opaque as a result of energy absorption via IB. The free-electron temperature increases faster than that of heavy-particles (up to more than $10^5$ K) \cite{DeMichelis_IEEE_1969,Zeldovich_book_1967,phuoc2002optical,alberti_JPhysD}, and a two lobes structure forms, with a rear lobe moving toward the laser source and a front lobe going in the opposite direction (\emph{backward} and \emph{forward} plasma waves, respectively) \cite{alberti_JPhysD,Tsuda_JApplPhys_1996,Tsuda_JApplPhys_2000}. It is worth recalling that the current model does not account for energy emission from the hot plasma kernel (that can be partly absorbed by the surrounding colder layer). Previous efforts from the authors \cite{alberti_JPhysD,alberti_thesis} have shown that the inclusion of energy emission and re-absorption by means of a diffusion approximation for radiative transfer is associated with a decrease of temperature and a radially larger plasma kernel. We omit these effects here for simplicity. After the laser-source is turned off, the plasma propagates within the cold surrounding gas, and in the first hundreds of nano-seconds the free-electron and heavy-particle temperatures equilibrate with each other and decrease down to \SI{10000}{\kelvin}--\SI{50000}{\kelvin} \cite{yalccin1999uence,el2009properties,limbach2014simultaneous,dumitrache2016threshold,harilal2015lifecycle,glumac2005temporal,alberti_dual}. With multi-mode beating, the breakdown dynamics is similar. However, as put in evidence by experiments \cite{nishihara2018influence}, the growth of the plasma kernel is accompanied by the development of local in-homogeneities. These are also observed in the current simulations as shown by the time-history of the mole fraction of free-electrons and their temperature in Fig. ~\ref{fig:multi_contour_fitted}, and by the evolution of the maximum value of the free-electron number density reported in Fig.~\ref{fig:multi_max_val_fitted}. A comparison with the laser time-history (see Fig. \ref{fig:multi_fitted}) indicates that the above local maxima are attained at the peaks of the laser power, suggesting the existence of a direct relation between the mode-beating frequency and the observed quasi-periodic structures. These matters are investigated in more detail in Sec.~\ref{sec:dynamics}. 
Figure~\ref{fig:multi_xt_diagram} reports the evolution of the plasma boundaries along the optical axis. For the simulations, the nominal plasma edge is taken to be the peak of the gradient of the mole fraction of free-electrons. Overall, the predicted evolution is in good agreement with the experiments ($3\sigma$ bounds are shown for the observations). The observed differences are thought to be primarily associated with the variability in the breakdown location, which Nishihara \emph{et al.} \cite{nishihara2018influence} quantify, for the multi-mode cases, to be of the order of \SI{0.4}{\milli\meter}. In addition, some parameters cannot be easily measured presently, which add a systematic uncertainty. In order to assess the effect of these uncertainties on the simulations, further calculations were conducted by changing the focal length up to \SI{20}{\percent} of the nominal value of \SI{100}{\milli\meter}, and by setting the focal radius equal to the upper-bound of \SI{10}{\micro\meter} determined by Nishihara \emph{et al.} via the razor edge method \cite{nishihara2018influence}. The results in terms of absorbed energy and evolution of the plasma boundaries along the optical axis are reported in Figs.~\ref{fig:multi_Eabs_fw} and \ref{fig:multi_xt_diagram_fw}, respectively. With the modified beam and optics parameters, the calculated plasma boundary evolution is in better agreement with the experiments. In analogy with the results of Fig. \ref{fig:multi_xt_diagram}, where nominal values are used, the agreement worsens with increasing energy. A similar trend is also observed when comparing the emission images as done in Fig. \ref{fig:multi_emission_fw} (details on the procedure to construct emission images from simulations may be found in Ref. \cite{alberti_JPhysD}).

\subsection{\label{sec:dynamics}Plasma formation mechanism: effect of mode-beating} 
The discussion in Sec.~\ref{sec:Validation} has shown that the LIB model is able to reproduce the two-lobed structure and its quasi-periodic spots observed in experiments for multi-mode pulses \cite{nishihara2018influence}. Here the effect of the mode-beating on plasma formation is investigated in detail.

\begin{figure}[htp]
\begin{center}
\includegraphics[width=0.5\textwidth]{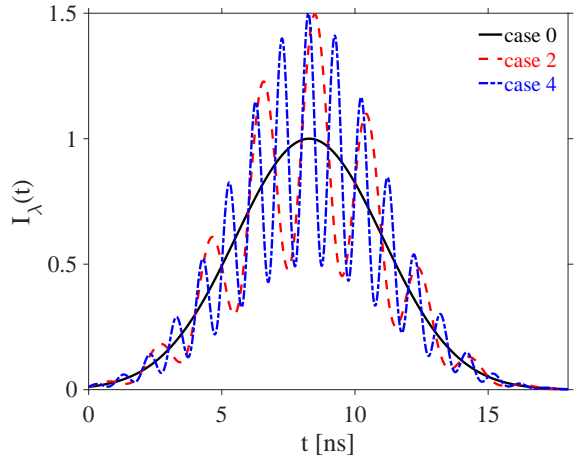}
\vspace*{-.2cm}
\caption{Temporal beam waveforms for cases 0 (Gaussian), 2 and 4.}
\label{fig:multi_analysis}
\end{center}
\end{figure}

In all simulations the incoming intensity distribution is evaluated based on Eq. \eqnref{eq:lasr_BC2} (no experimental data used) for the purpose of studying the effect of the mode-beating frequency on the breakdown dynamics. A total of seven cases (0--6) is here considered. Case 0 is a purely Gaussian pulse (\emph{i.e.}, $A_{mb} = 0$), whereas cases 1 through 6 represent multi-mode pulses with non-dimensional amplitude $A_{mb} = \num{0.5}$, and mode-beating frequencies $f_{mb}$ equal to \SIlist{250;500;750;1000;1500;2000}{\mega\hertz}. For all cases, the temporal FWHM, the laser energy and the focal length are \SI{6.5}{\nano\second}, \SI{60}{\milli\joule} and \SI{75}{\milli\meter}, respectively. All the other parameters (\emph{e.g.}, ambient conditions, focal radius) are the same of the baseline values of Sec. \ref{sec:Validation}. The temporal beam waveforms for cases 0 (Gaussian), 2 and 4 are plotted in Fig.~\ref{fig:multi_analysis}.

\begin{figure}[htp]
\begin{center}
\includegraphics[width=0.5\textwidth]{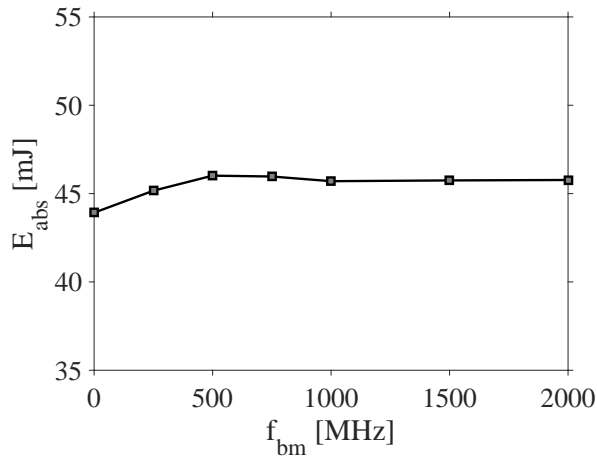}
\vspace*{-.2cm}
\caption{Absorbed energy for cases 0 (Gaussian) through 6.}
\label{fig:multi_Eabs_analysis}
\end{center}
\end{figure}

Figure \ref{fig:multi_Eabs_analysis} displays the absorbed energy for cases 0 through 6. The plotted data show an increase of the absorbed energy up to frequencies $\approx \SI{500}{\mega\hertz}$. For larger values, the absorbed energy undergoes a small decrease before approaching a flat profile. Energy deposition is maximum for case 2, though the increase with respect to the single-mode (\emph{i.e.}, Gaussian) case is modest ($\approx 2-3 \%$). The above results may be explained based on the fact that, during the time intervals when the mode-beating function is $g_{mb}(t) > 1$, the intensity is larger than that of the corresponding Gaussian pulse (see also Fig. \ref{fig:multi_analysis}). Since the volumetric rate of energy deposition via IB is proportional to the intensity (\emph{i.e.}, $\Omega = \kappa I$), this leads to larger absorbed energies. The effect becomes more and more negligible beyond a threshold frequency, since higher frequencies shorten the time-interval during which $g_{mb}(t) > 1$.  
\begin{figure*}[htp]
\begin{center}
\includegraphics[width=0.5\textwidth]{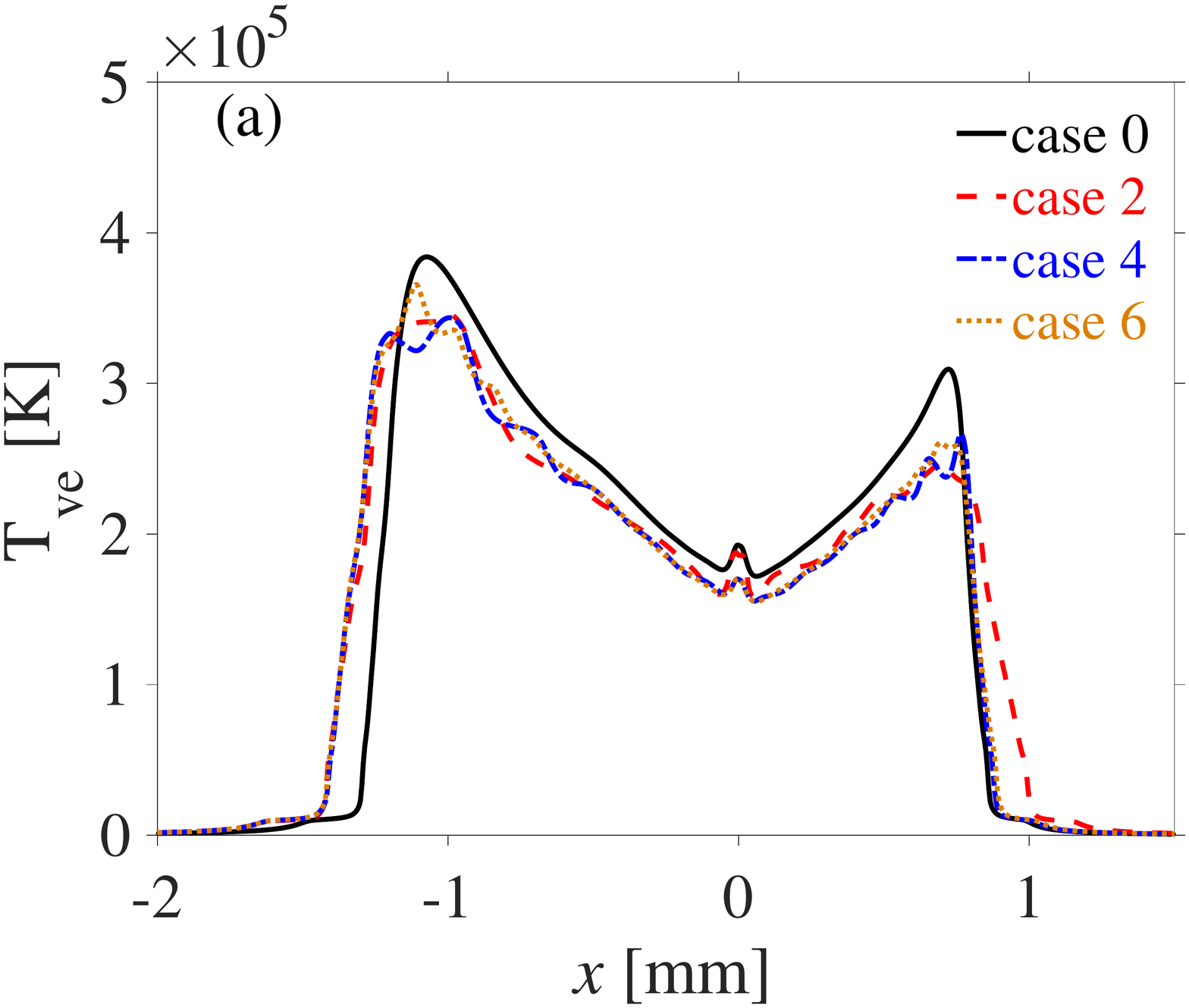}
\hspace{-0.5cm}
\includegraphics[width=0.5\textwidth]{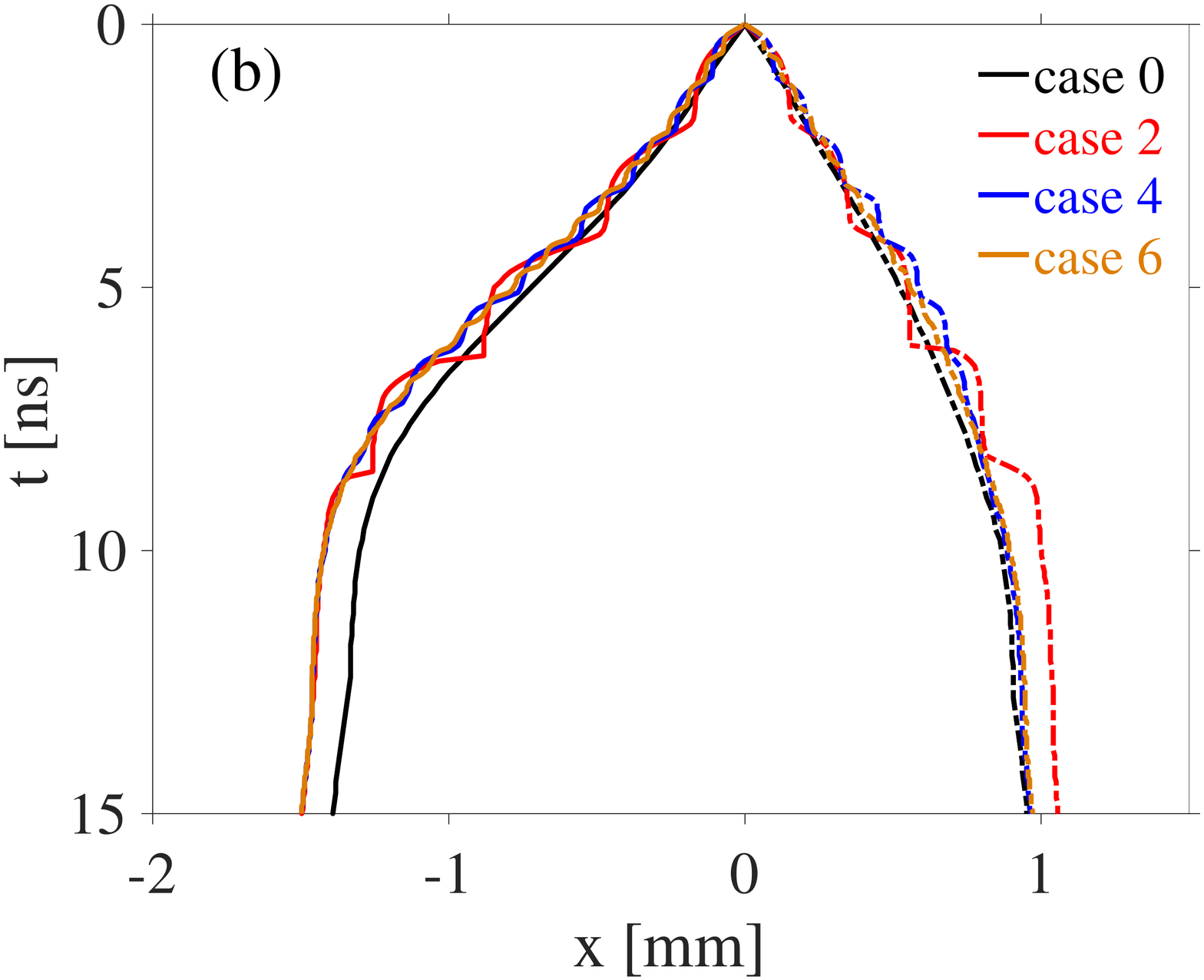}
\vspace*{-.2cm}
\caption{(a) Free-electron temperature along the optical axis for cases 0 (Gaussian), 2, 4 and 6 at $t = \SI{10}{\nano\second}$, and (b) $x$-$t$ diagram of plasma boundary evolution along the optical axis (solid lines are rear lobe and broken lines are front lobe).}
\label{fig:multi_analysis_sliceX_1}
\end{center}
\end{figure*}
Changing the mode-beating frequency affects also the growth of the plasma kernel, as demonstrated by the free-electron temperature profiles and $x$-$t$ diagrams in Fig.~\ref{fig:multi_analysis_sliceX_1}. In general, the multi-mode pulses lead to a more elongated plasma compared to the single-mode case. This finding is consistent with the above discussion on absorbed energy. Among the multi-mode results, case 2 leads to a larger front lobe than cases 4 and 6, whereas the differences are barely noticeable for the rear lobe. This is a consequence of the fact that, at the rear lobe, energy deposition is essentially dependent only on the instantaneous value of the intensity, which is the reason why, toward the end of the pulse, all multi-mode plasma boundary profiles collapse onto the same curve. On the other hand, energy deposition at the front lobe is affected by beam attenuation which, in turn, depends not only on the instantaneous optical properties (\emph{e.g.}, absorption coefficient) but also on the time-history of the pulse, since the latter has a influence on the evolution of the plasma kernel.

\begin{figure}[htp]
\begin{center}
\includegraphics[width=0.5\textwidth]{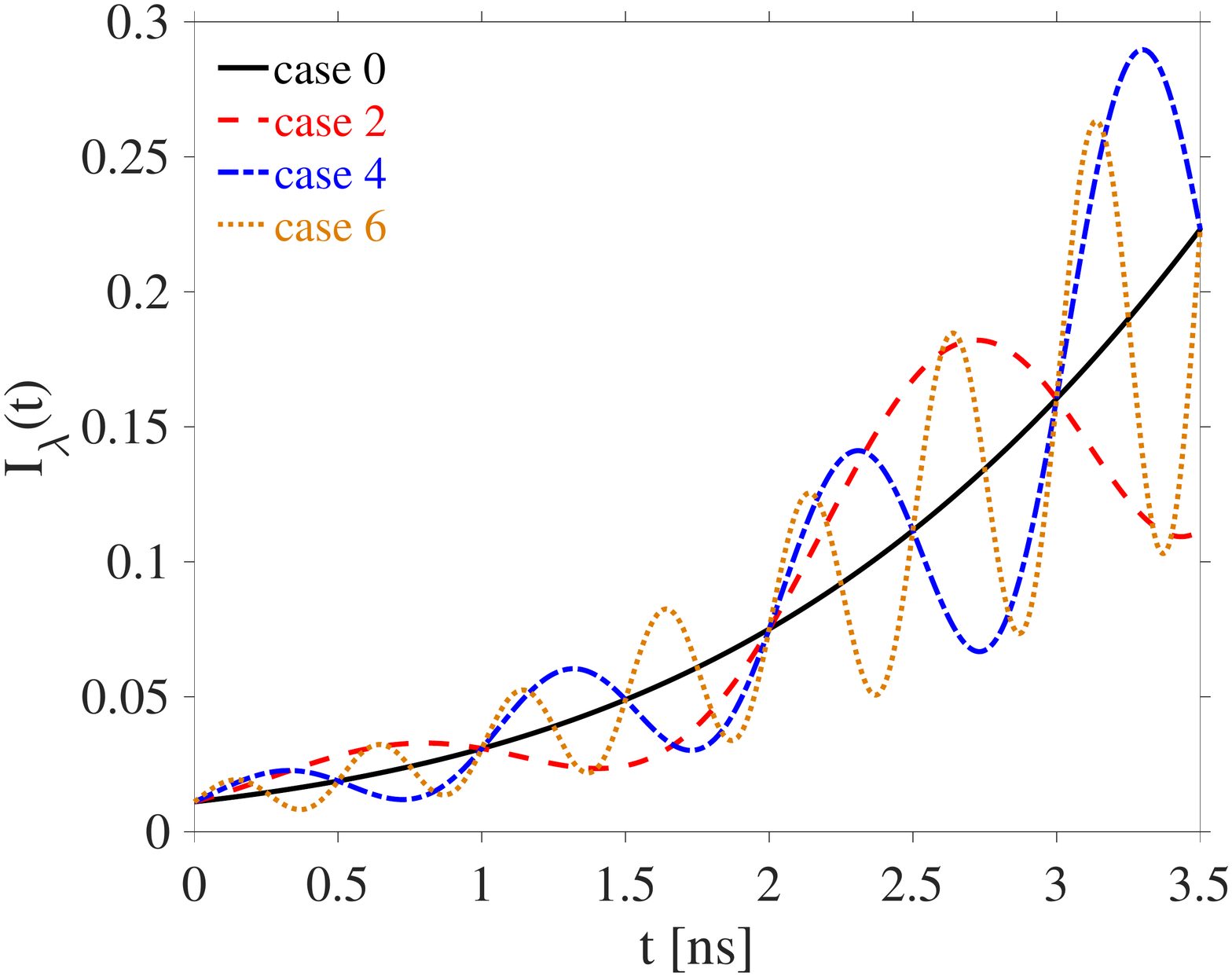}
\vspace*{-.2cm}
\caption{Temporal beam waveform for cases 0 (Gaussian), 2, 4 and 6 in the time interval $0 \leq t \leq \SI{3.5}{\nano\second}$.}
\label{fig:multi_analysis_temporal_zoom}
\end{center}
\end{figure}

\begin{figure*}
\begin{center}
\includegraphics[width=0.5\textwidth]{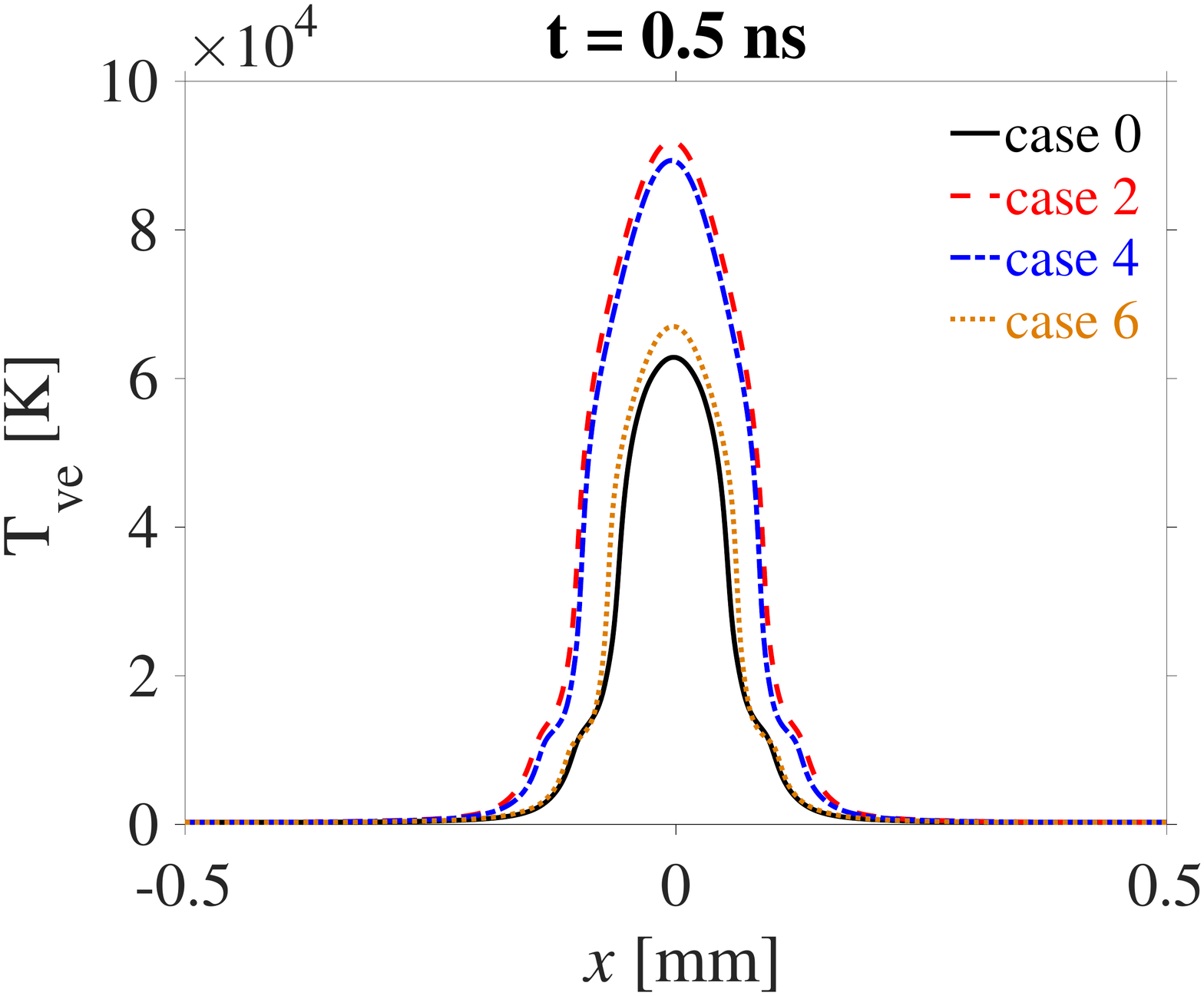}
\hspace{-0.5cm}
\includegraphics[width=0.5\textwidth]{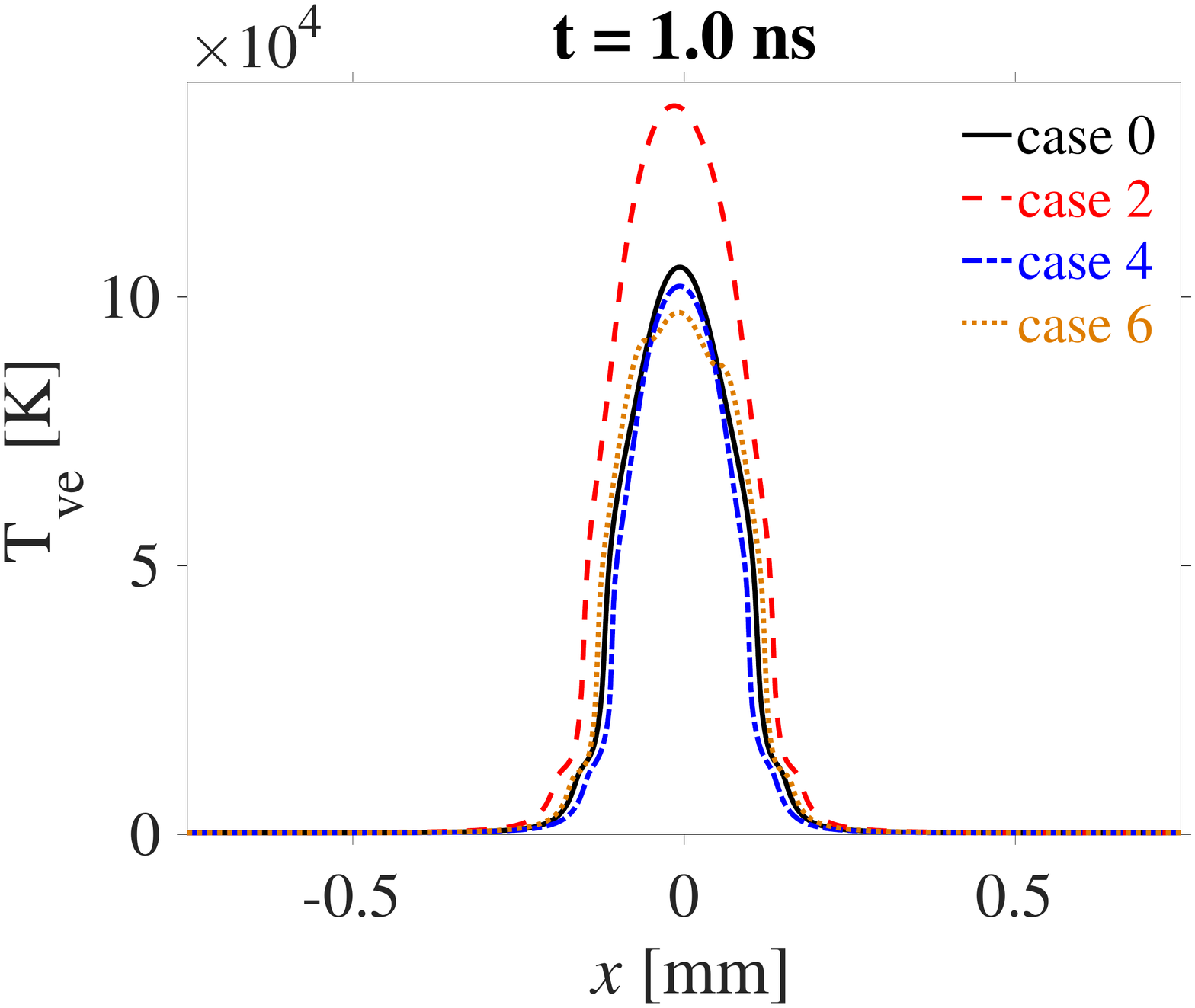} 
\includegraphics[width=0.5\textwidth]{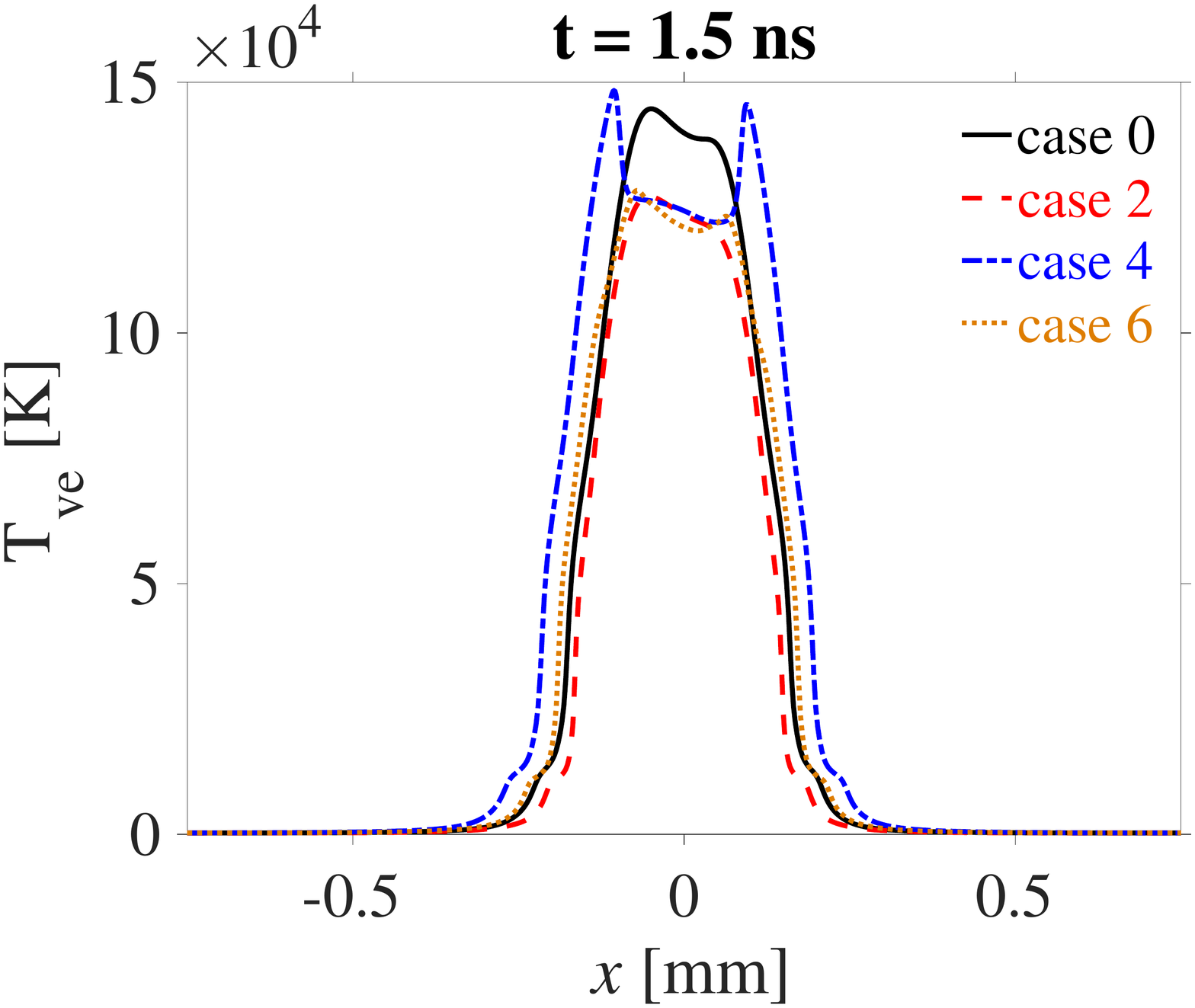}
\hspace{-0.5cm}
\includegraphics[width=0.5\textwidth]{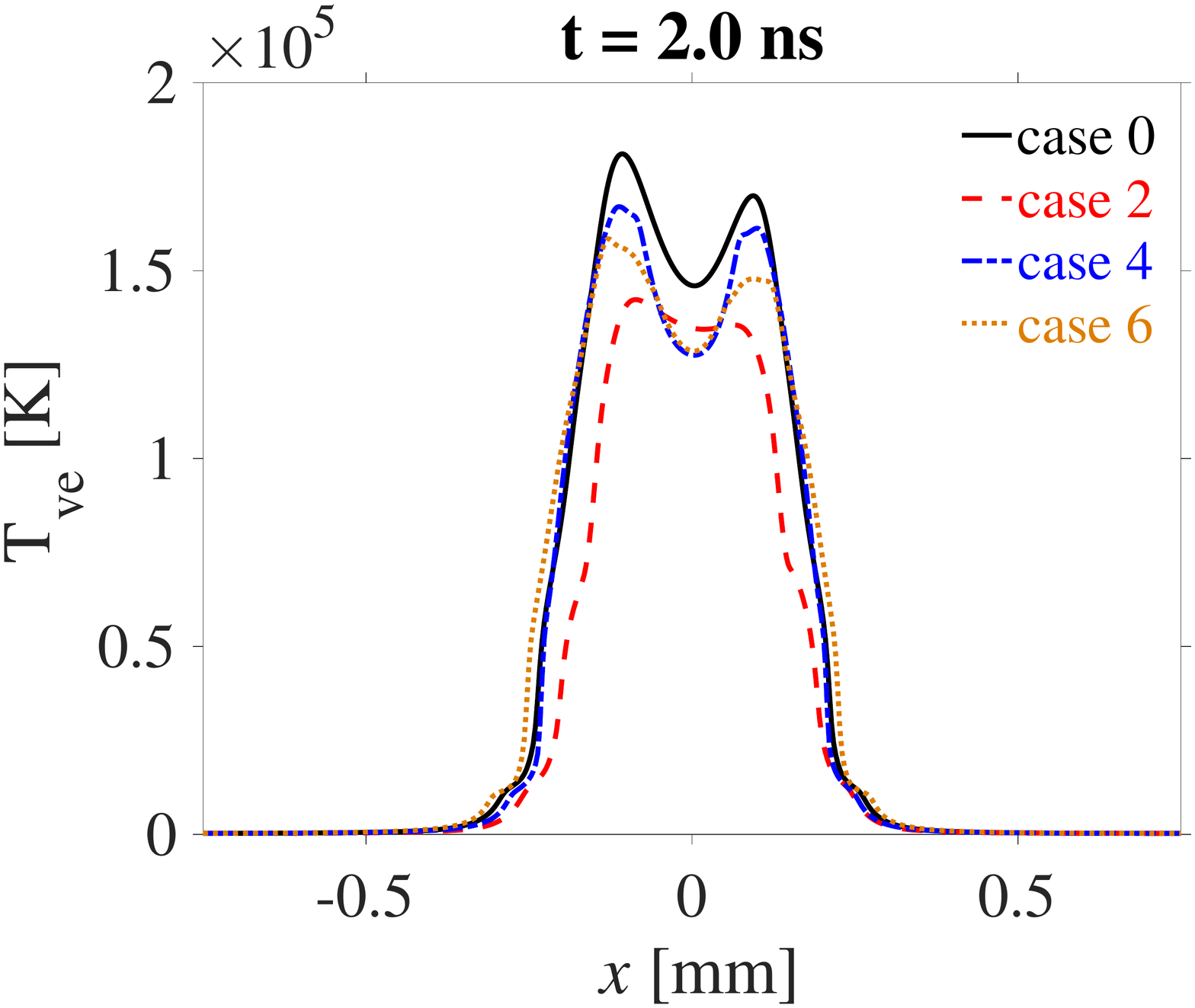} 
\includegraphics[width=0.5\textwidth]{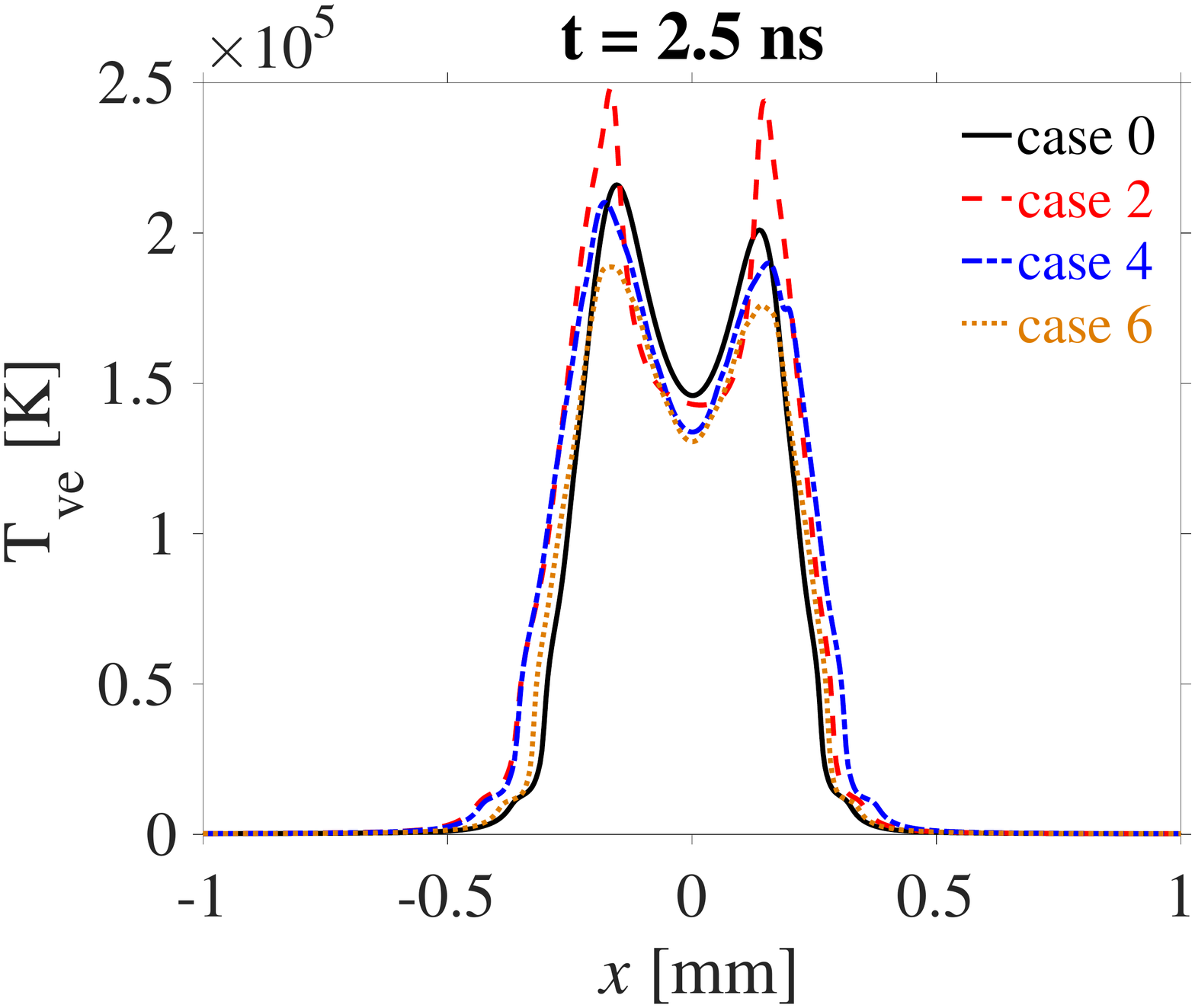}
\hspace{-0.5cm}
\includegraphics[width=0.5\textwidth]{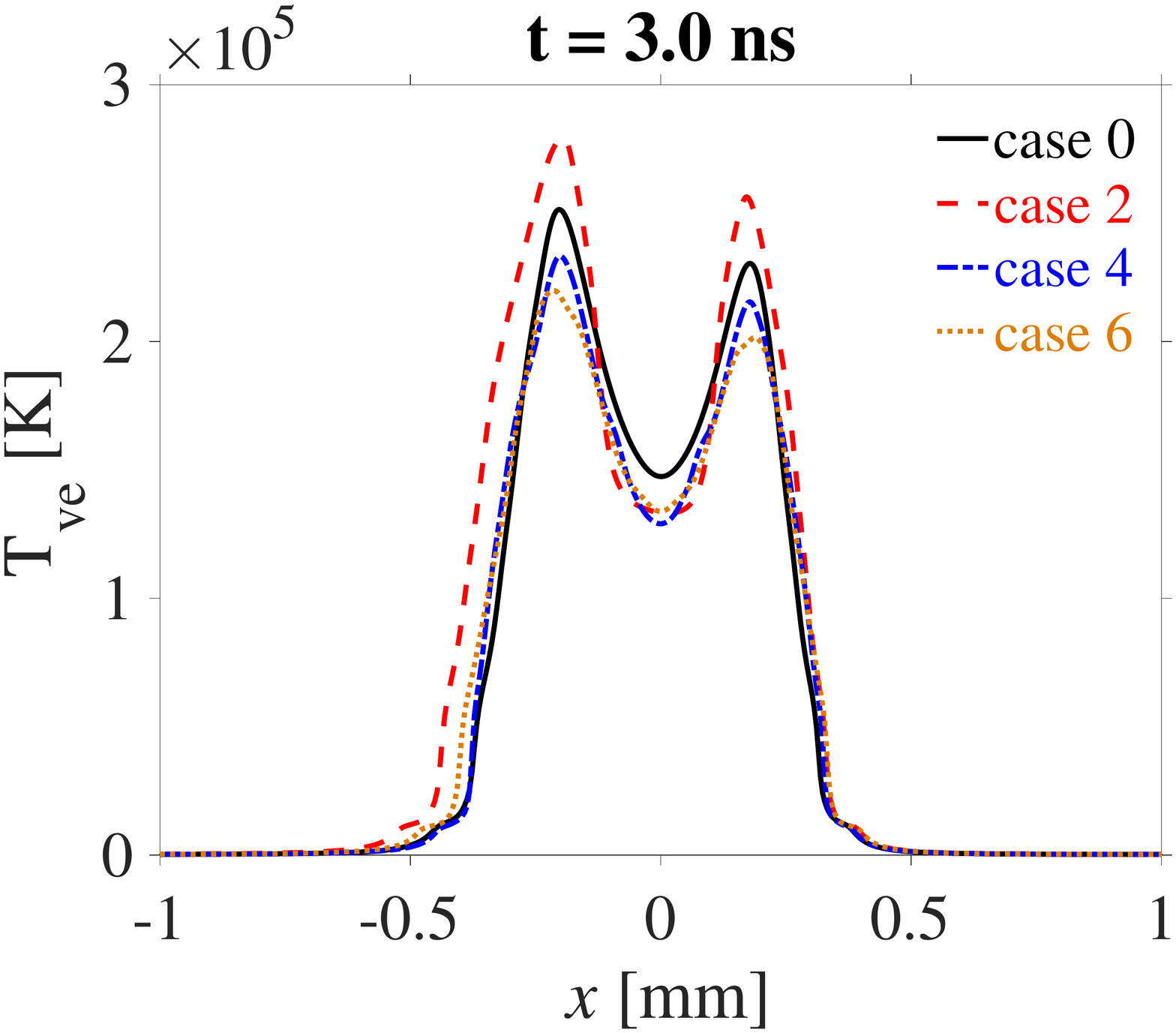} 
\vspace*{-0.2cm}
\caption{Temporal evolution of the free-electron temperature along the optical axis for cases 0 (Gaussian), 2, 4 and 6 in the time interval $0 \leq t \leq \SI{3.5}{\nano\second}$.}
\label{fig:multi_analysis_sliceX_2_early_times}
\end{center}
\end{figure*}

The impact of the beam temporal waveform on the plasma dynamics may be observed when monitoring the free-electron temperature. This is done in Fig. \ref{fig:multi_analysis_sliceX_2_early_times} showing the evolution of the latter quantity along the optical axis at the onset of breakdown (\emph{i.e.}, $\num{0} \leq t \leq \SI{3.5}{\nano\second}$). For the sake of convenience, the beam waveform for the considered time-frame is also reported in Fig. \ref{fig:multi_analysis_temporal_zoom}. During $0 \leq t \leq \SI{1}{ns}$, the intensity is larger for case 2 and the associated free-electron temperature rises faster than cases 4 and 6, as shown by the frames extracted at $t = \SI{0.5}{ns}$ and $t = \SI{1.0}{ns}$. The same behavior is observed for cases 4 and 6 during the intervals $\num{1} \leq t \leq \SI{1.5}{ns}$ and $\num{1.5} \leq t \leq \SI{2.0}{ns}$, respectively, which correspond to the regions where the related mode-beating functions attain local maxima. Proceeding further in time, another power-peak phase for case 2 is entered, leading to a new sharp rise of the free-electron temperature around $t = \SI{2.5}{\nano\second}$. Overall, the amplitude-modulation of a baseline Gaussian pulse alters the heating rate of the plasma, compared to the single-mode case, as revealed by the observed sudden rises of the free-electron temperature.

\begin{figure*}[htp]
\begin{center}
\includegraphics[width=0.5\textwidth]{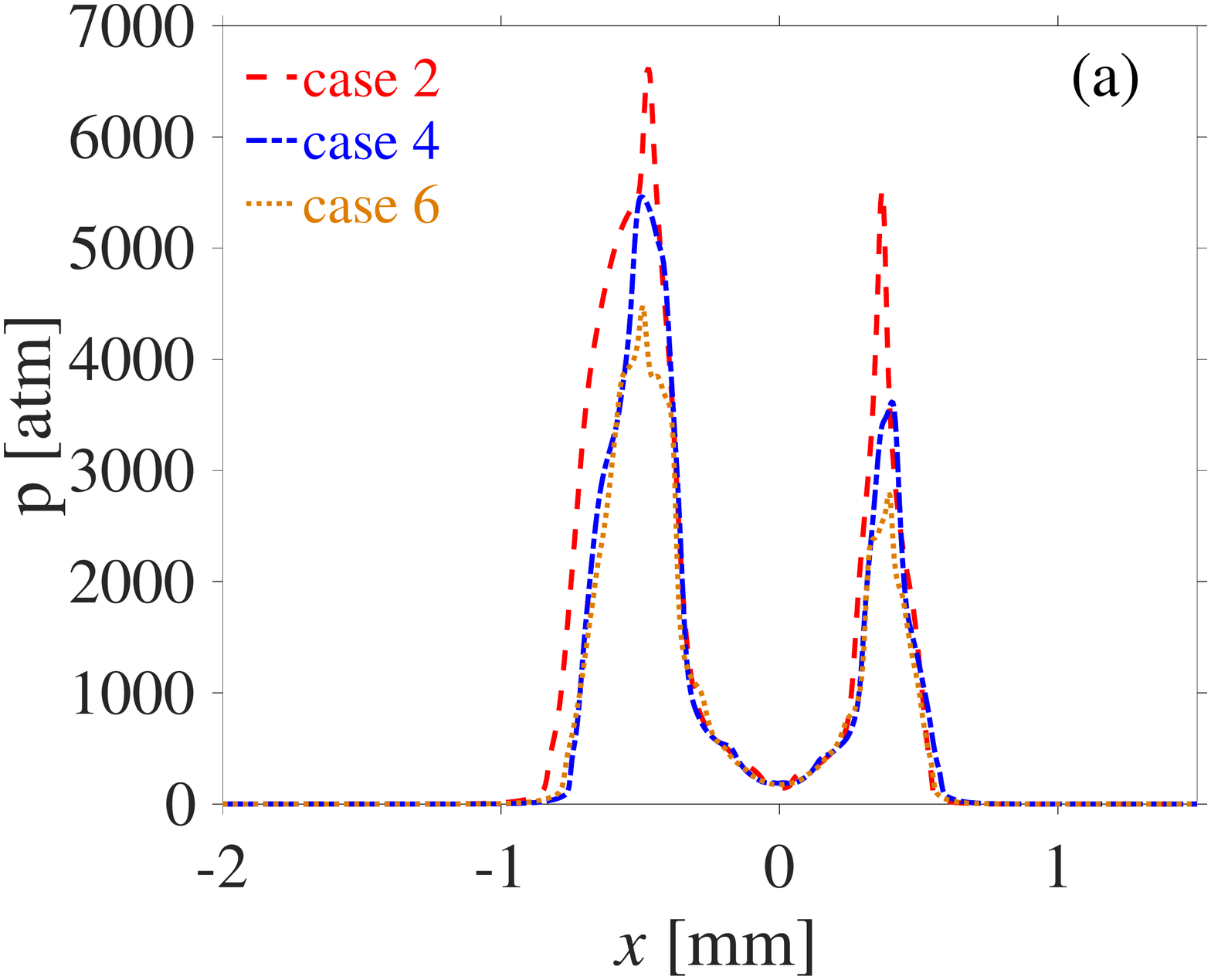}
\hspace{-0.5cm}
\includegraphics[width=0.5\textwidth]{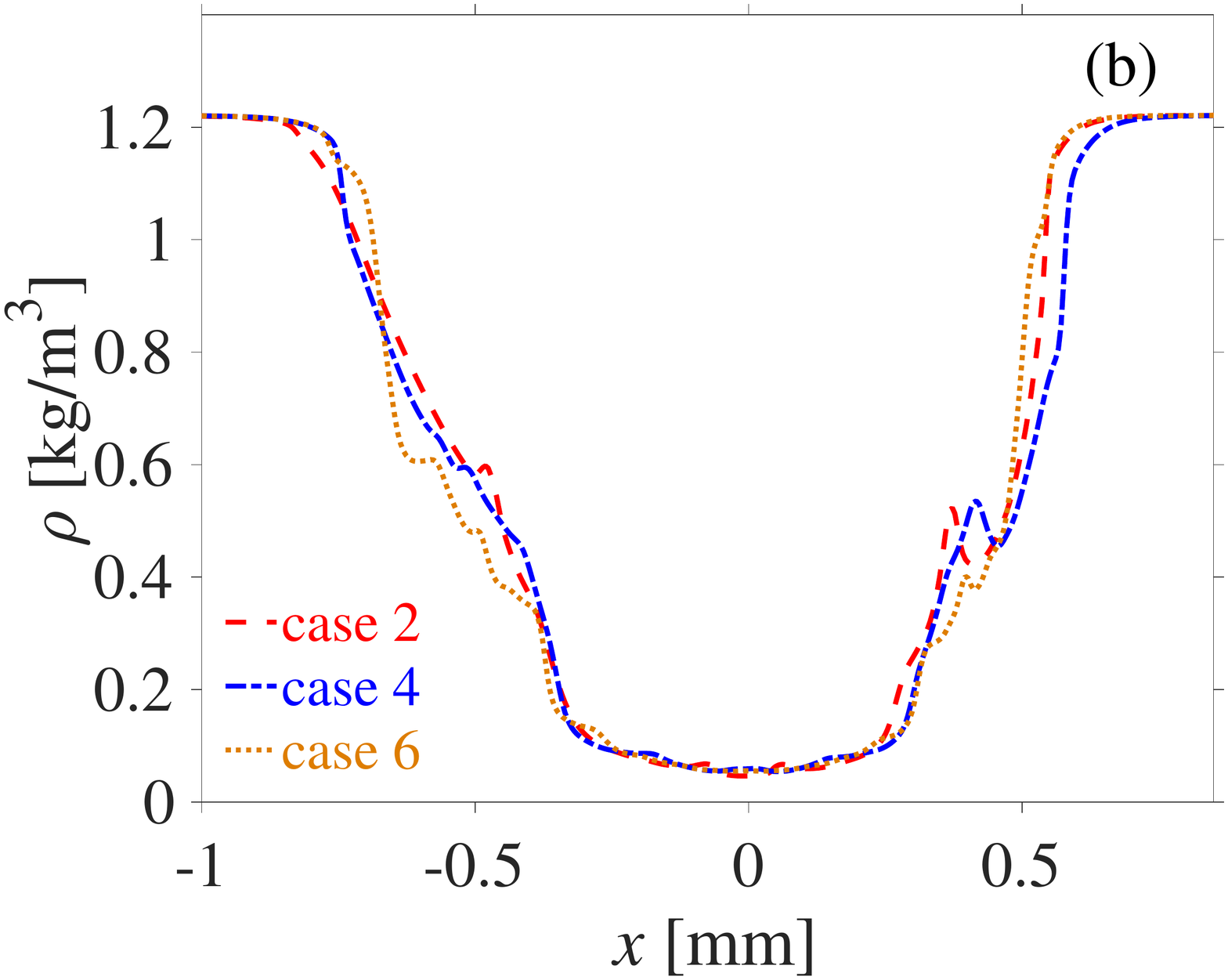}
\vspace*{-.2cm}
\caption{Distribution of pressure (a) and density (b) along the optical axis at $t = \SI{5}{\nano\second}$.}
\label{fig:multi_analysis_sliceXY_5ns}
\end{center}
\end{figure*}

\begin{figure}[htp]
\begin{center}
\includegraphics[width=0.5\textwidth]{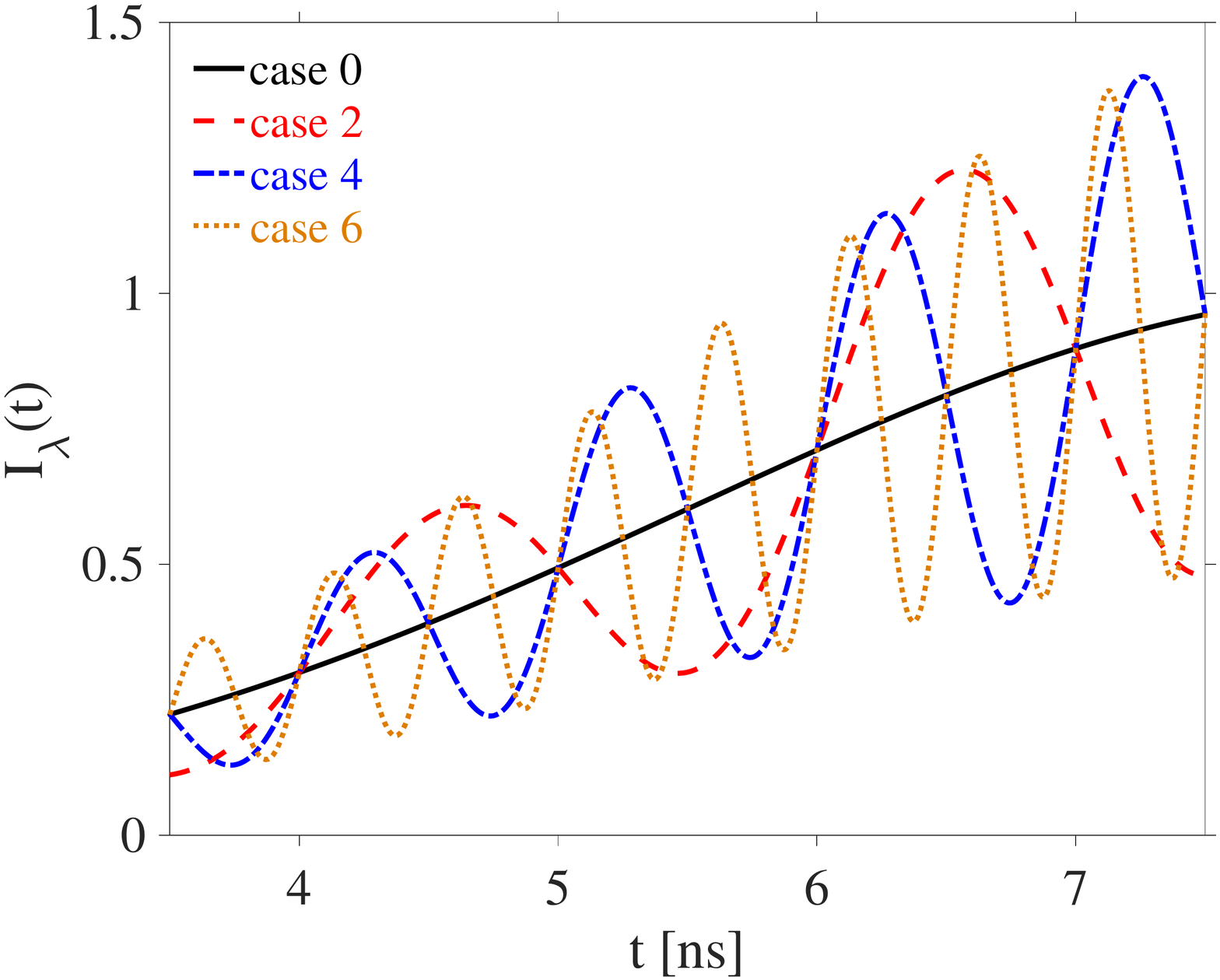}
\vspace*{-.2cm}
\caption{Temporal beam waveform for cases 0 (Gaussian), 2, 4 and 6 in the time interval $\num{3.5} \leq t \leq \SI{7.5}{ns}$.}
\label{fig:multi_analysis_temporal_zoom2}
\end{center}
\end{figure}

\begin{figure*}
\begin{center}
\includegraphics[width=0.5\textwidth]{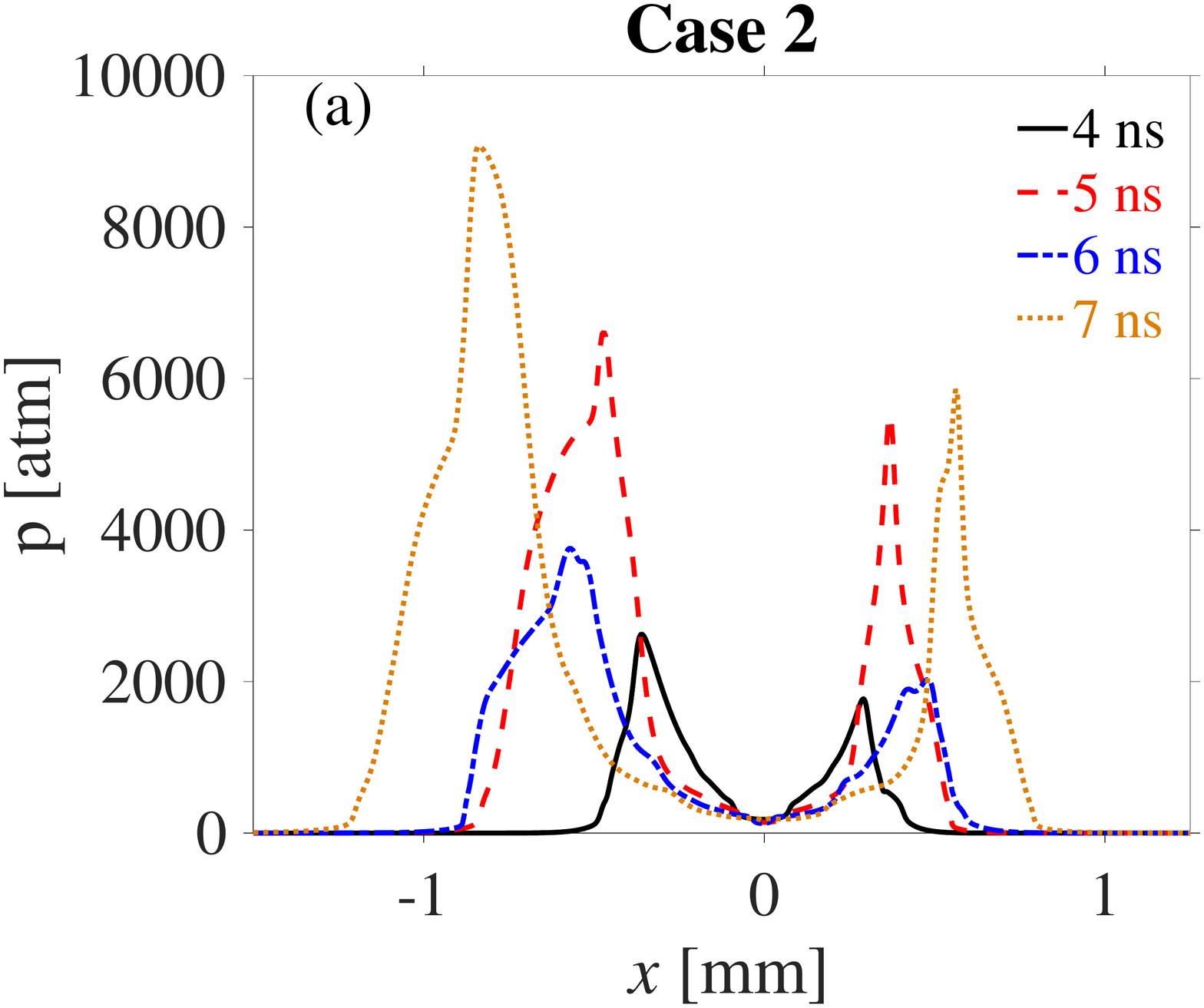}
\hspace{-0.5cm}
\includegraphics[width=0.5\textwidth]{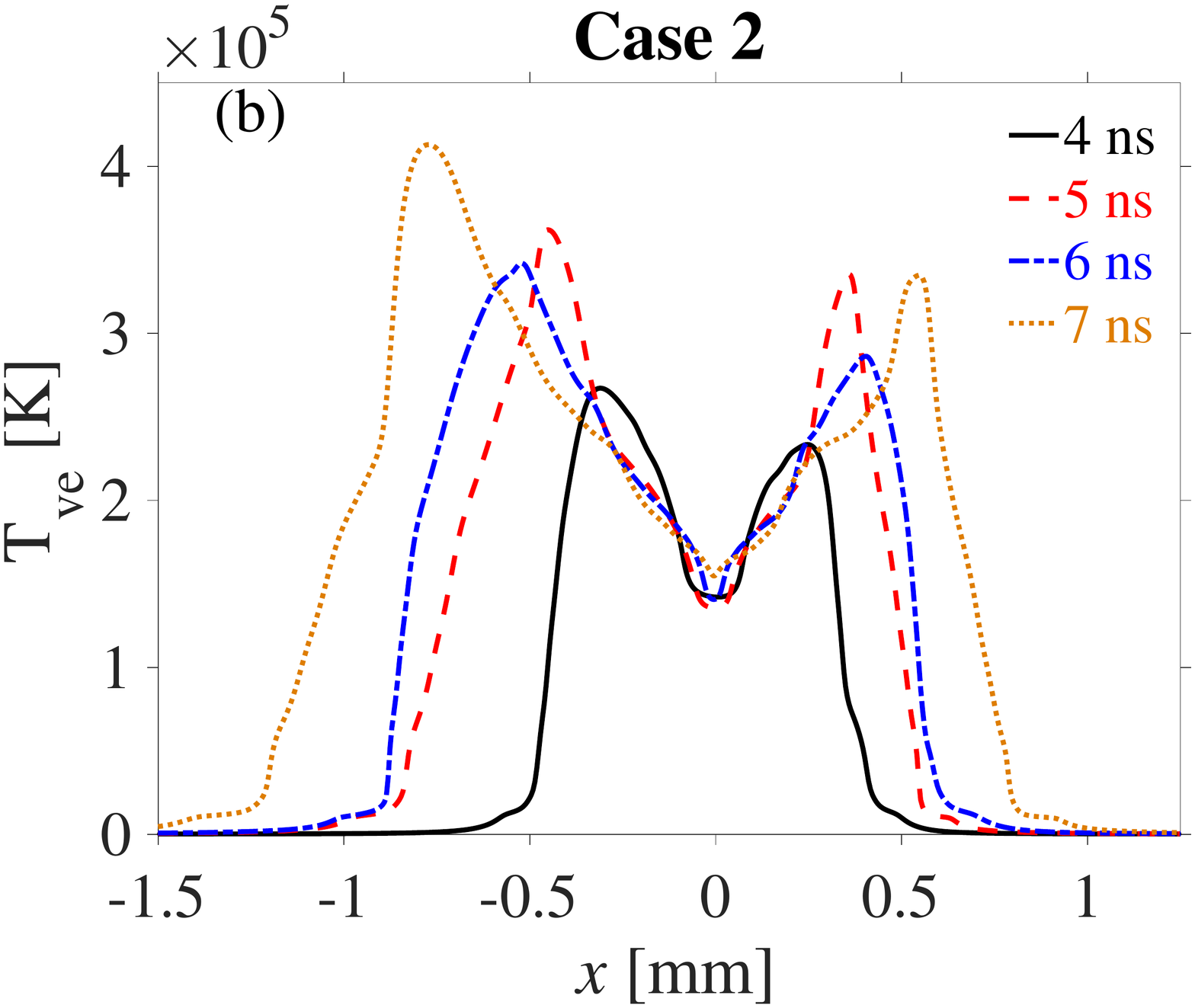} 
\includegraphics[width=0.5\textwidth]{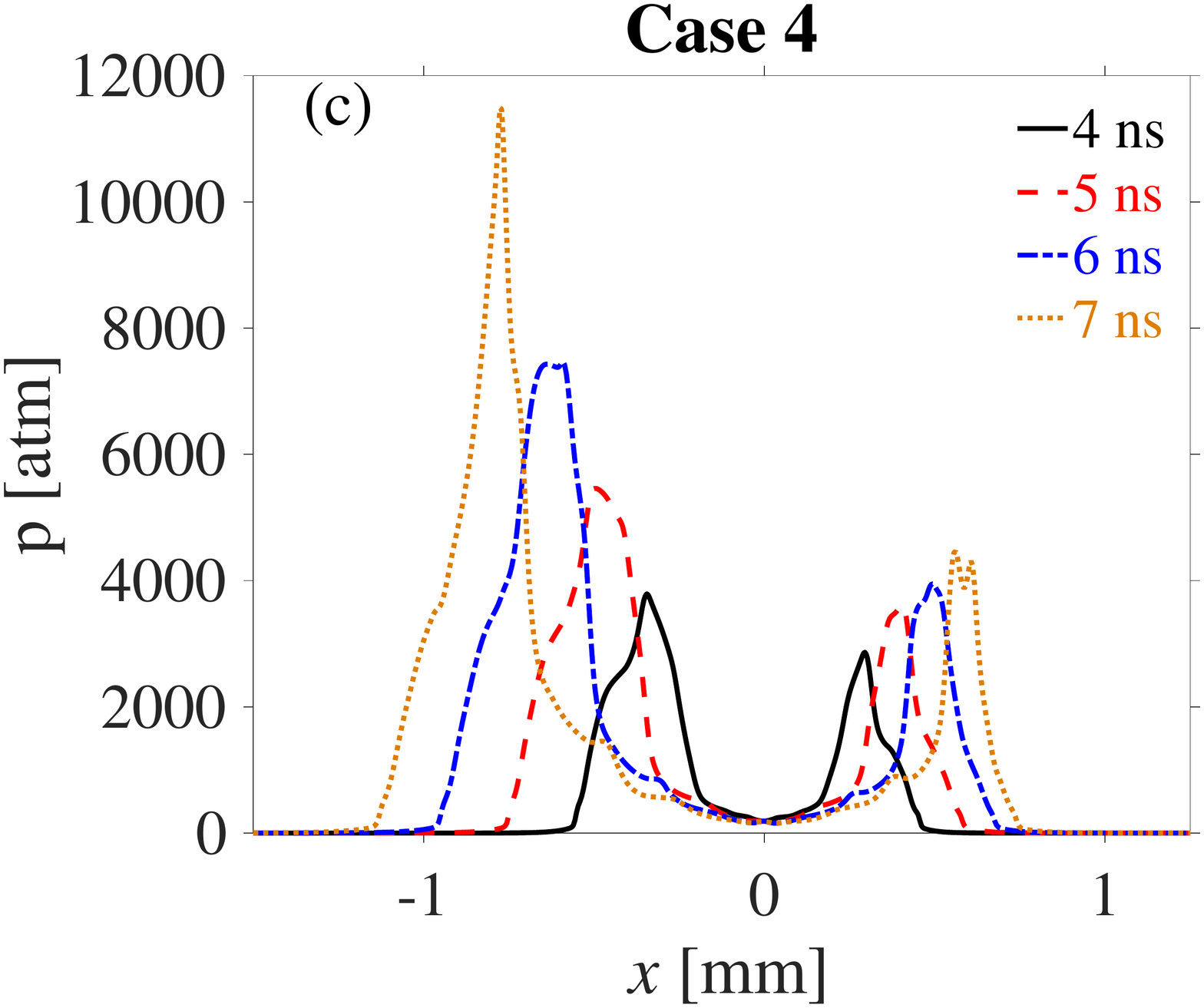}
\hspace{-0.5cm}
\includegraphics[width=0.5\textwidth]{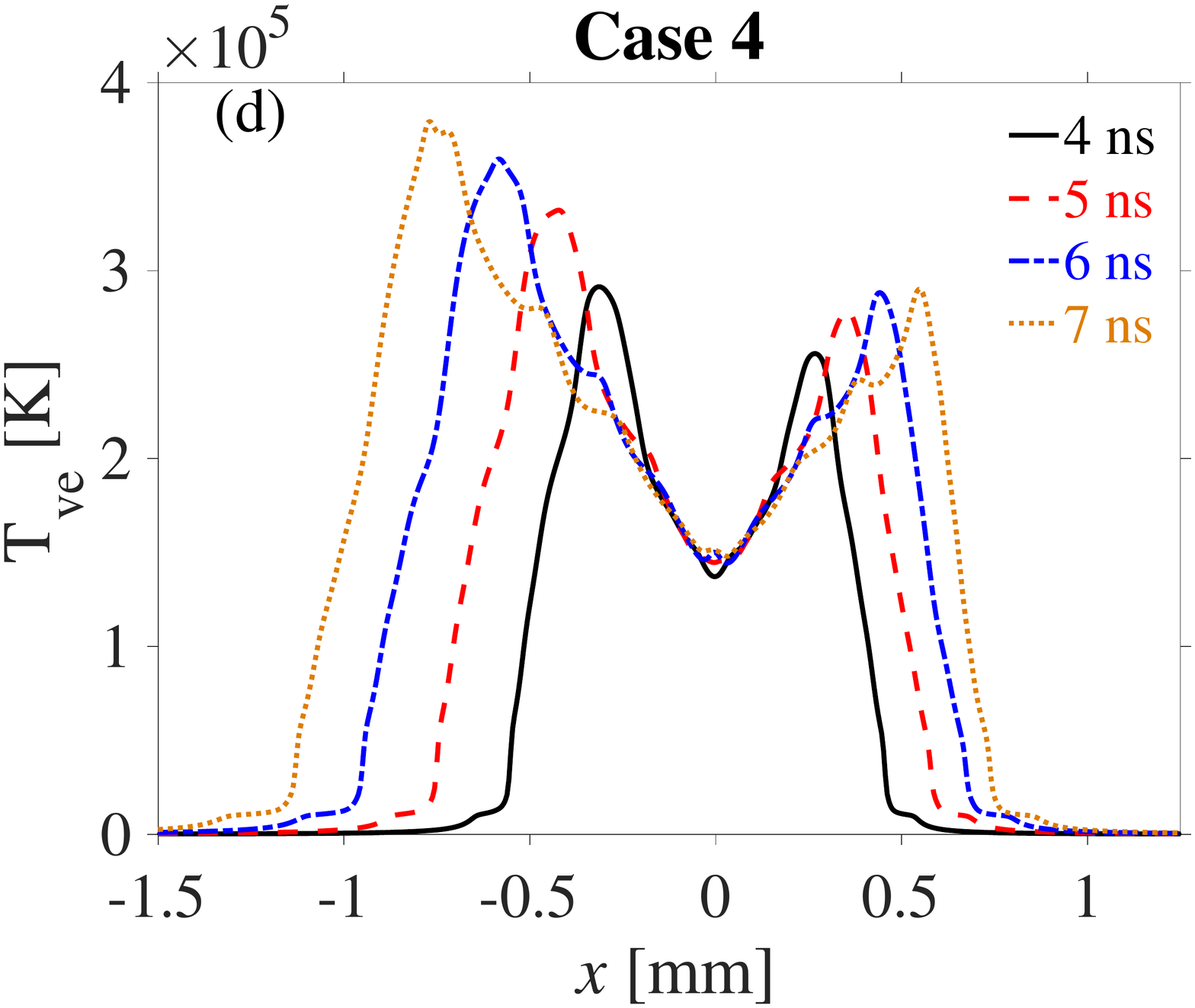} 
\includegraphics[width=0.5\textwidth]{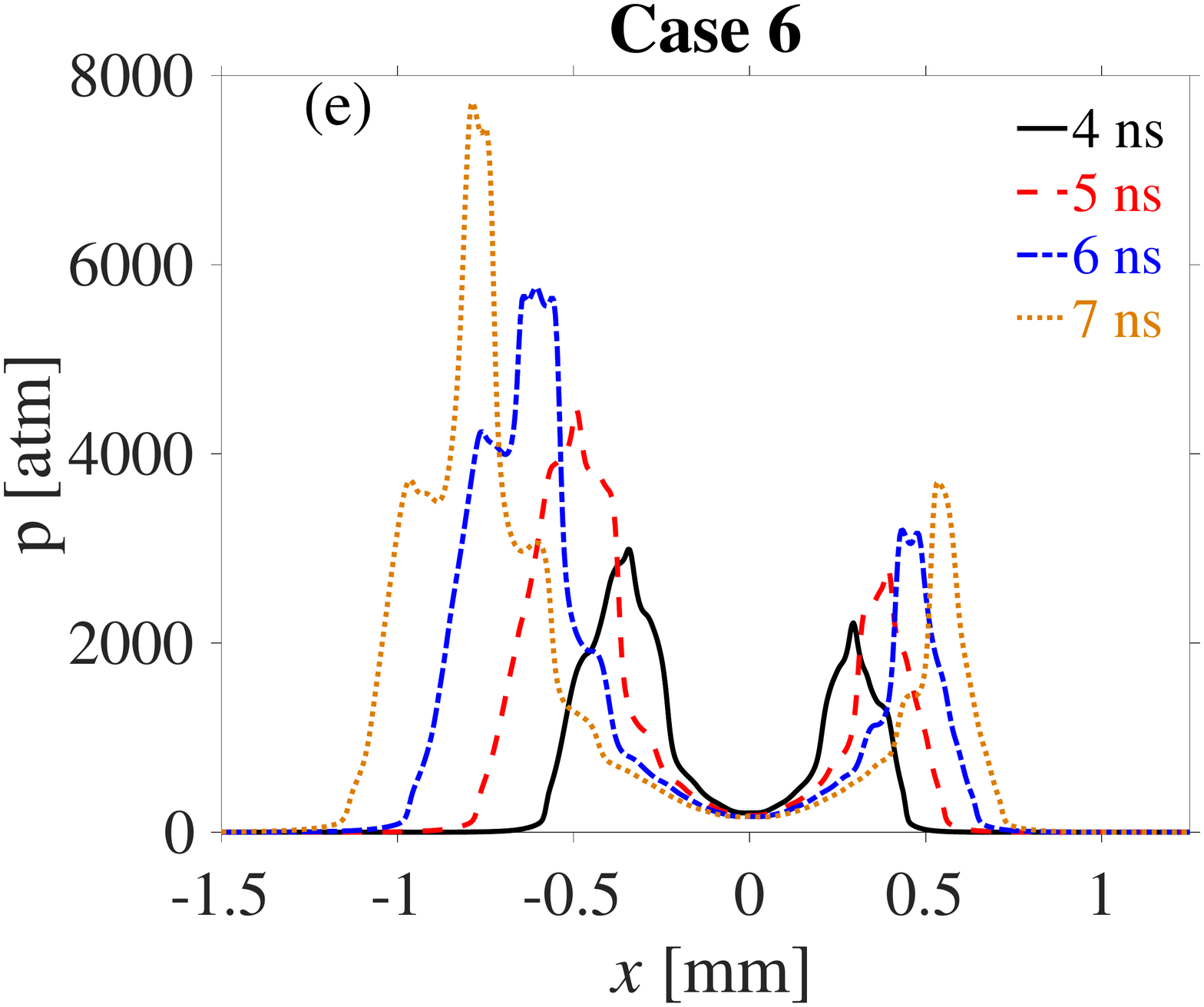}
\hspace{-0.5cm}
\includegraphics[width=0.5\textwidth]{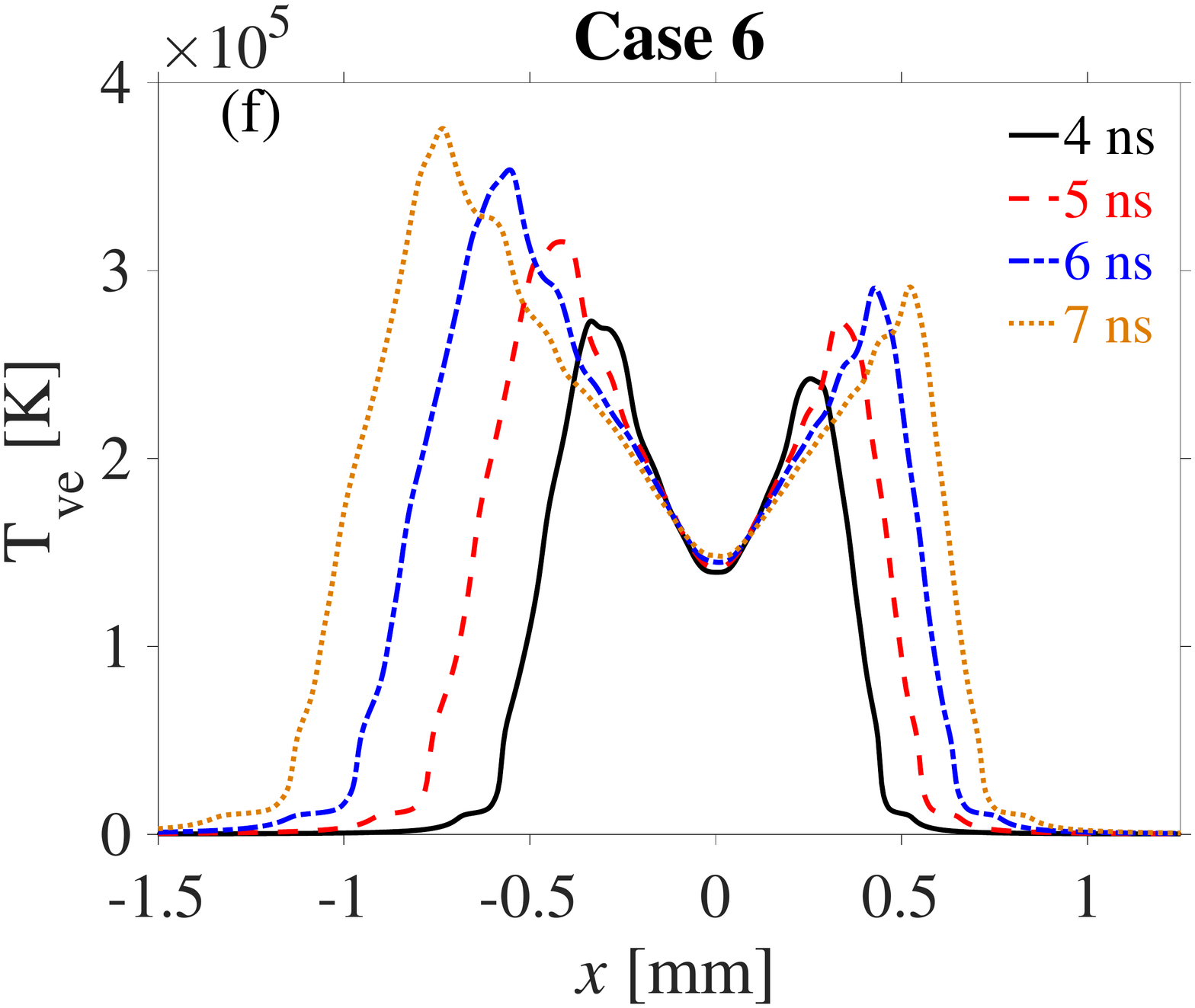} 
\vspace*{-.2cm}
\caption{Temporal evolution of the distribution of pressure (a,c,e) and free-electron temperature (b,d,f) along the optical axis.}
\label{fig:multi_analysis_sliceX_3_later_times}
\end{center}
\end{figure*}

\begin{figure*}[htp]
\begin{center}
\includegraphics[width=1.0\textwidth]{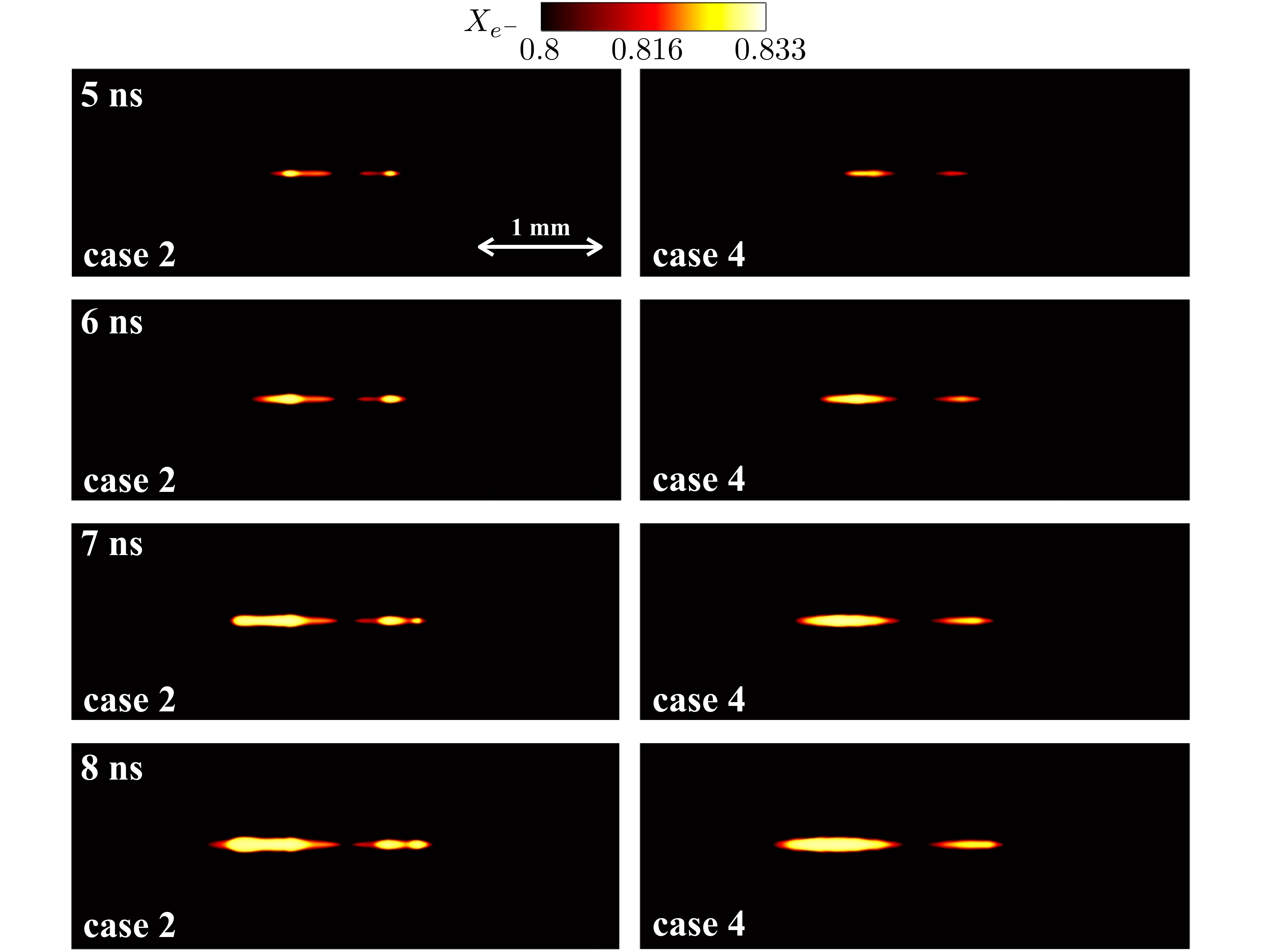}
	\caption{Temporal evolution of the mole fraction of free-electrons for cases 2 and 4 highlighting local in-homogeneities resulting from the multi-mode operating fashion.}
\label{fig:multi_contour_analysis}
\end{center}
\end{figure*}

At $t \approx \SI{5}{ns}$, the large increase of pressure associated with the absorbed laser energy (see Fig.~\ref{fig:multi_analysis_sliceXY_5ns}(a)) generates a strong gas-dynamic expansion in the plasma core, as can be observed from the density distribution reported in Fig.~\ref{fig:multi_analysis_sliceXY_5ns}(b). The sudden expansion generates a reduction of the gas density and, consequently, a decrease of the absorption coefficient. This, together with the low power portion of the laser temporal history for case~2 during the interval $\num{5} \leq t \leq \SI{6}{ns}$ (see Fig.~\ref{fig:multi_analysis_temporal_zoom2}), prevents the free-electron temperature from rising again (\emph{i.e.}, the sudden expansion compensates for the IB energy absorption), as it can be observed in Fig.~\ref{fig:multi_analysis_sliceX_3_later_times} (top right for $\SI{5}{ns}$ and $\SI{6}{ns}$). In turn, a larger fraction of the laser radiation is able to  escape the plasma rear lobe and be absorbed by the front lobe. The same does not occur for cases~4 and 6 since, in the same time interval, at least one new peak of the laser intensity temporal history deposits enough energy to sustain the growth of the rear lobe as shown in Fig.~\ref{fig:multi_analysis_sliceX_3_later_times}, with the consequence that radiation attenuation precludes the laser energy from being efficiently deposited in the front lobe at later times. The overall effect on plasma formation is shown in Fig. \ref{fig:multi_contour_analysis} which compares the evolution of the mole fraction of free-electrons of cases 2 and 4. Increasing the mode-beating frequency leads to a more homogeneous plasma resembling the tear-drop morphology of a single-mode pulse \cite{alberti_JPhysD}. 

\begin{figure*}[htp]
\begin{center}
\includegraphics[width=1.0\textwidth]{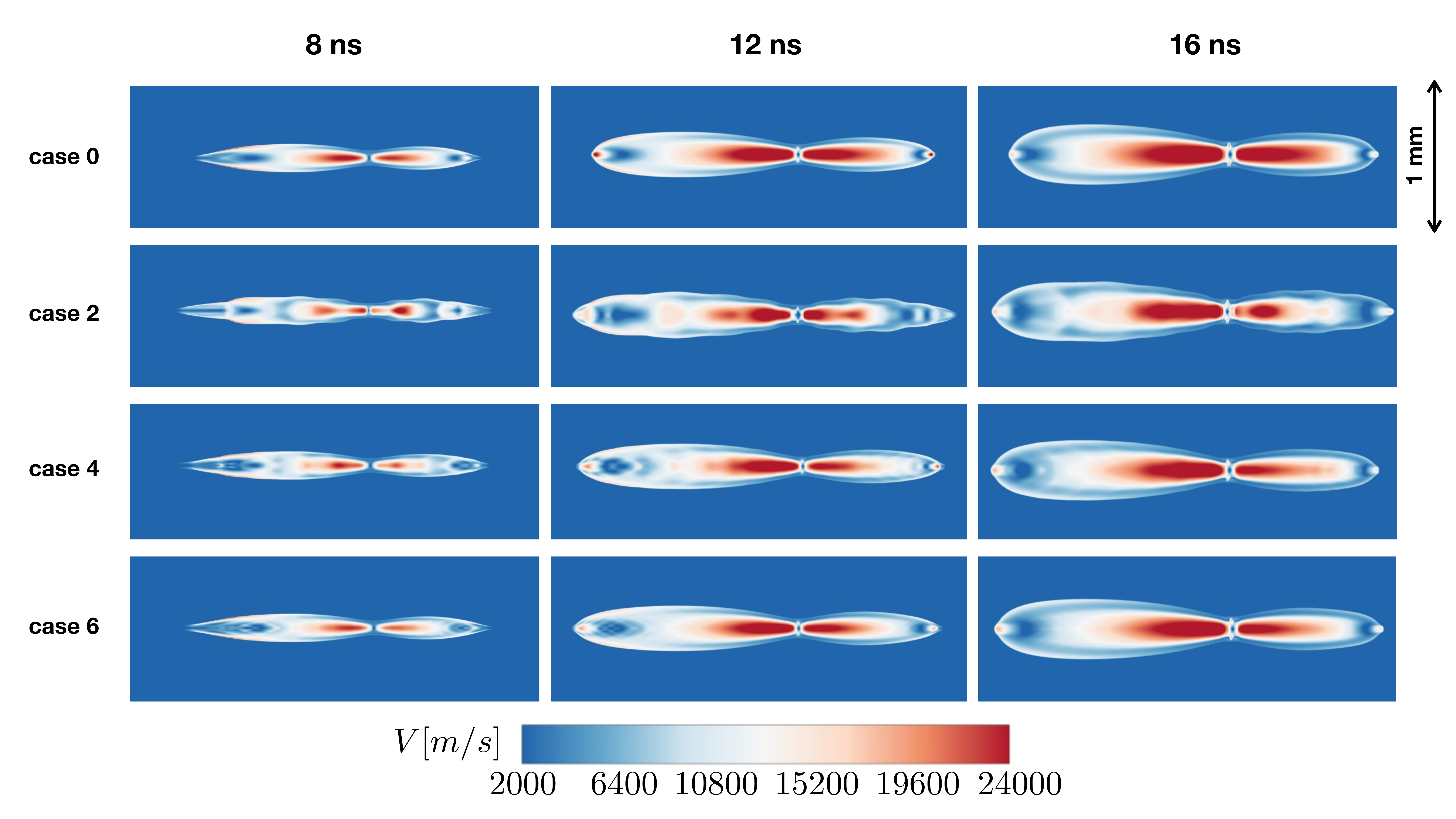}
\vspace*{-.2cm}
\caption{Temporal evolution of the velocity field during the laser discharge.}
\label{fig:velocity_fields_single_multi}
\end{center}
\end{figure*}

Fig.~\ref{fig:velocity_fields_single_multi} shows the temporal evolution of the velocity magnitude at $\SI{8}{ns}$, $\SI{12}{ns}$, and $\SI{16}{ns}$ for cases 0 (Gaussian), 2, 4, and 6. Mode-beating results in more irregular velocity profiles, however no overall amplification is observed compared to case 0 (\emph{i.e.}, Gaussian pulse). The velocity contours are analogous for all cases, suggesting that for the conditions investigated in this work hydrodynamic instability is not responsible for the onset of the plasma kernels, since there is no evidence of amplifying instabilities associated with the inflectional velocity profiles.

\begin{figure*}[htp]
\begin{center}
\includegraphics[width=0.35\textwidth]{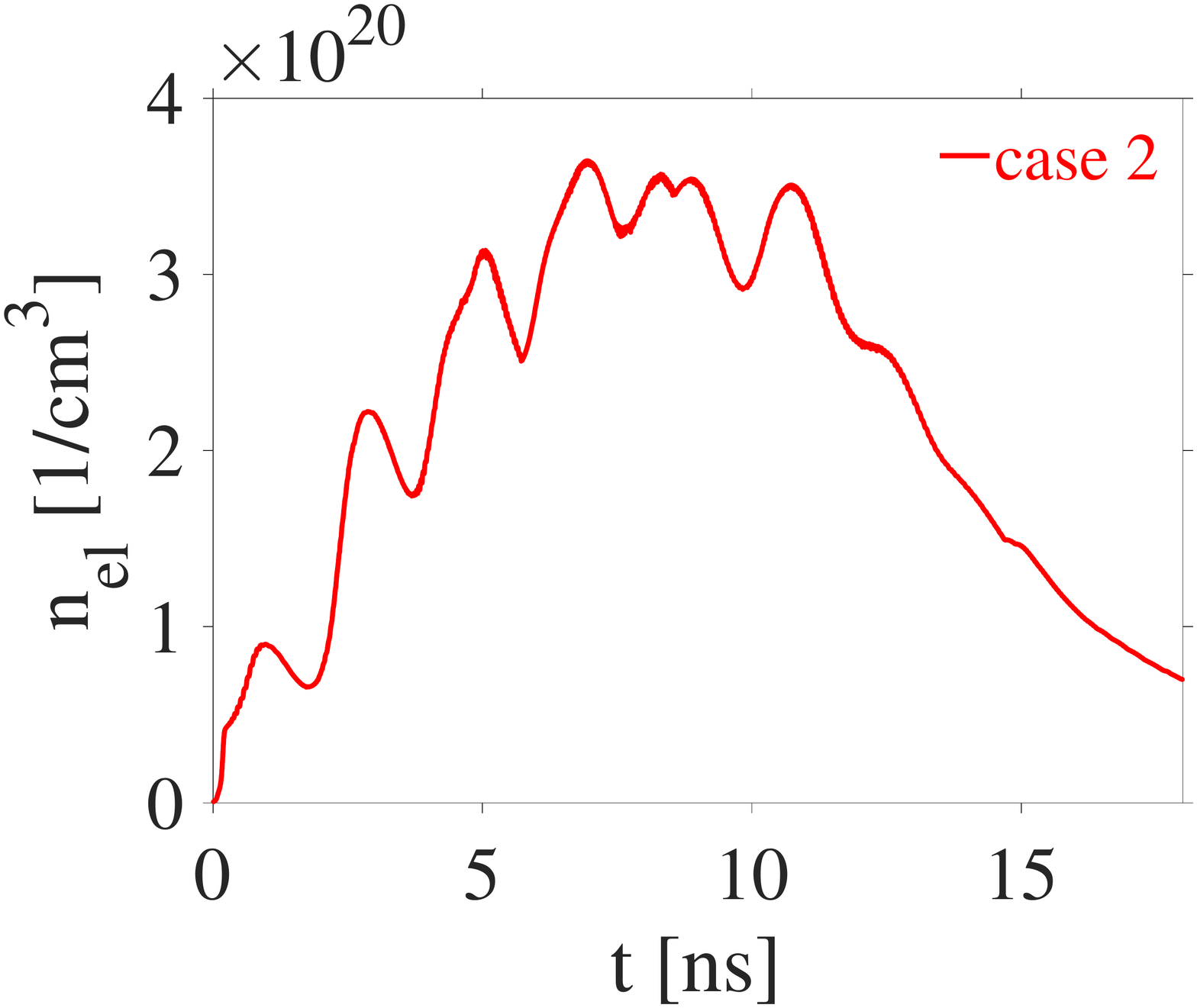}
\hspace{-0.7cm}
\includegraphics[width=0.35\textwidth]{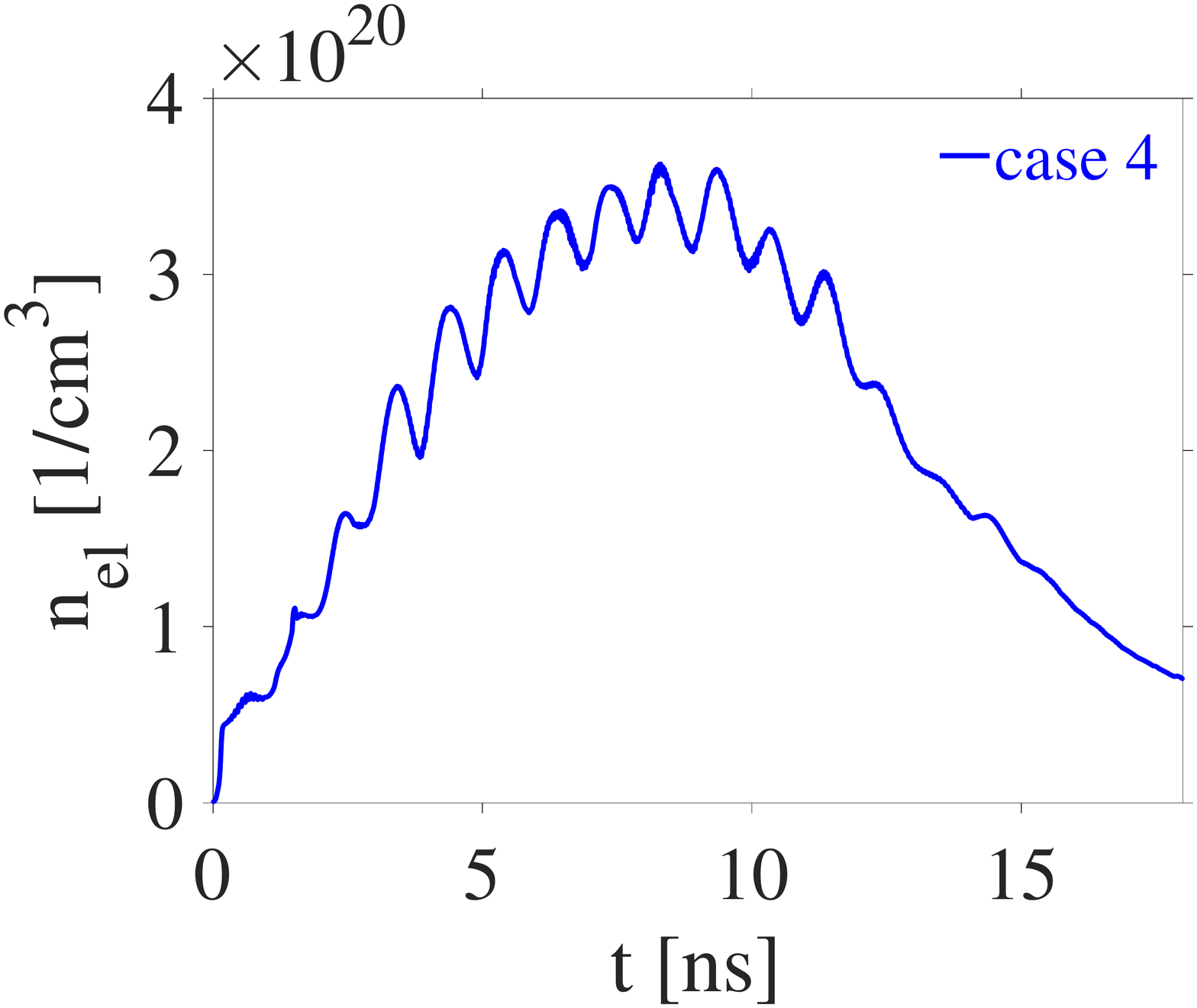}
\hspace{-0.5cm}
\includegraphics[width=0.35\textwidth]{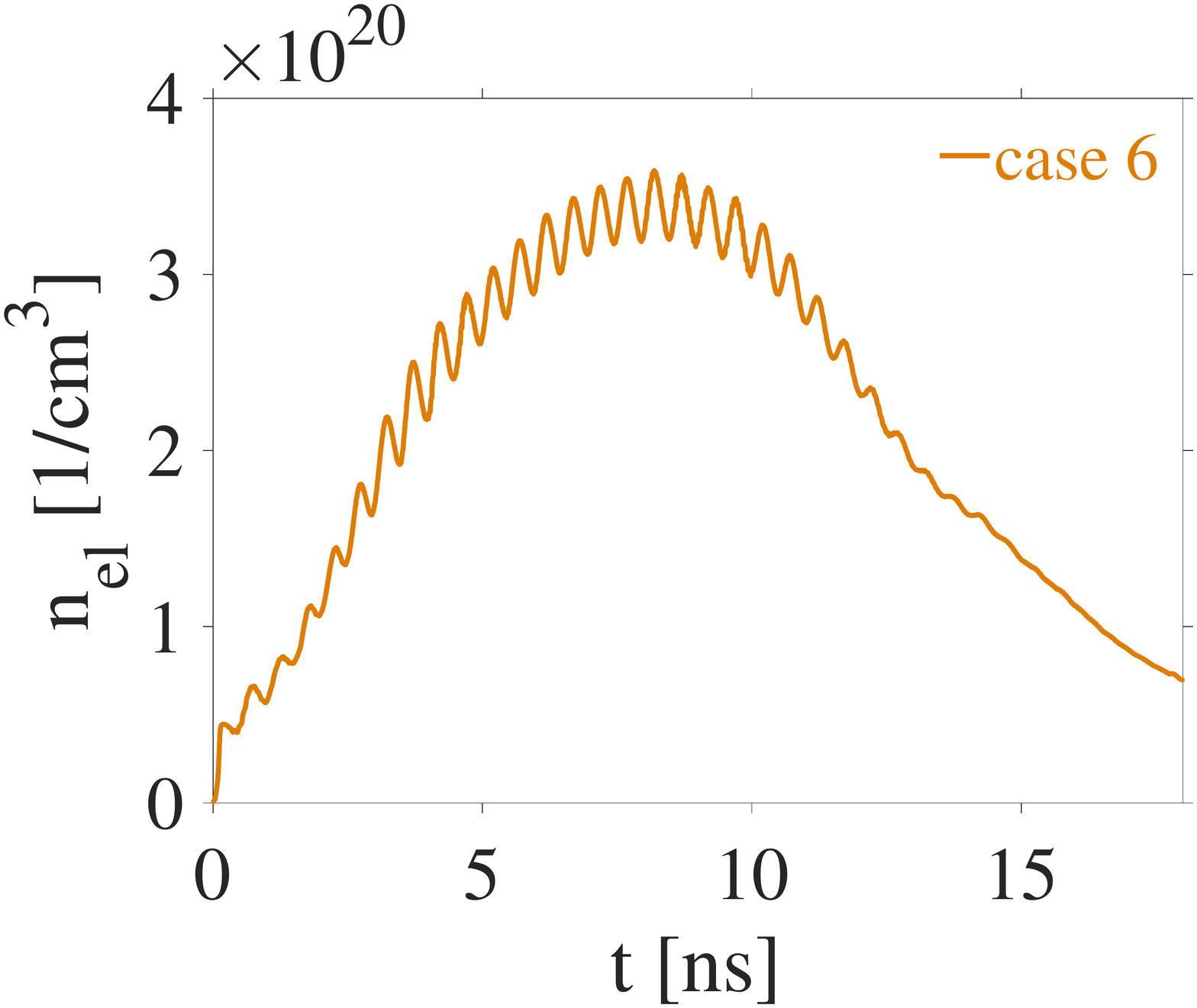}
\includegraphics[width=0.35\textwidth]{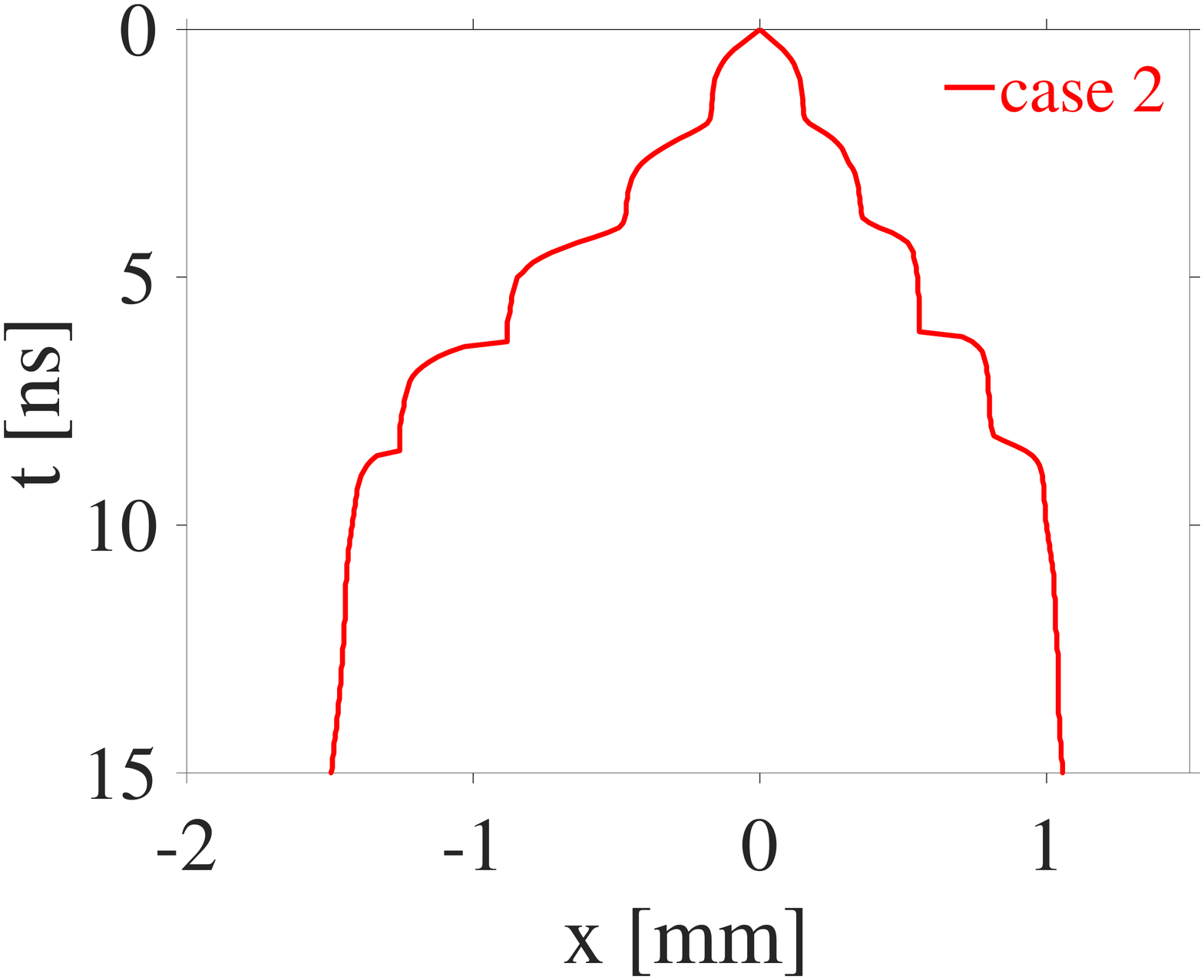}
\hspace{-0.7cm}
\includegraphics[width=0.35\textwidth]{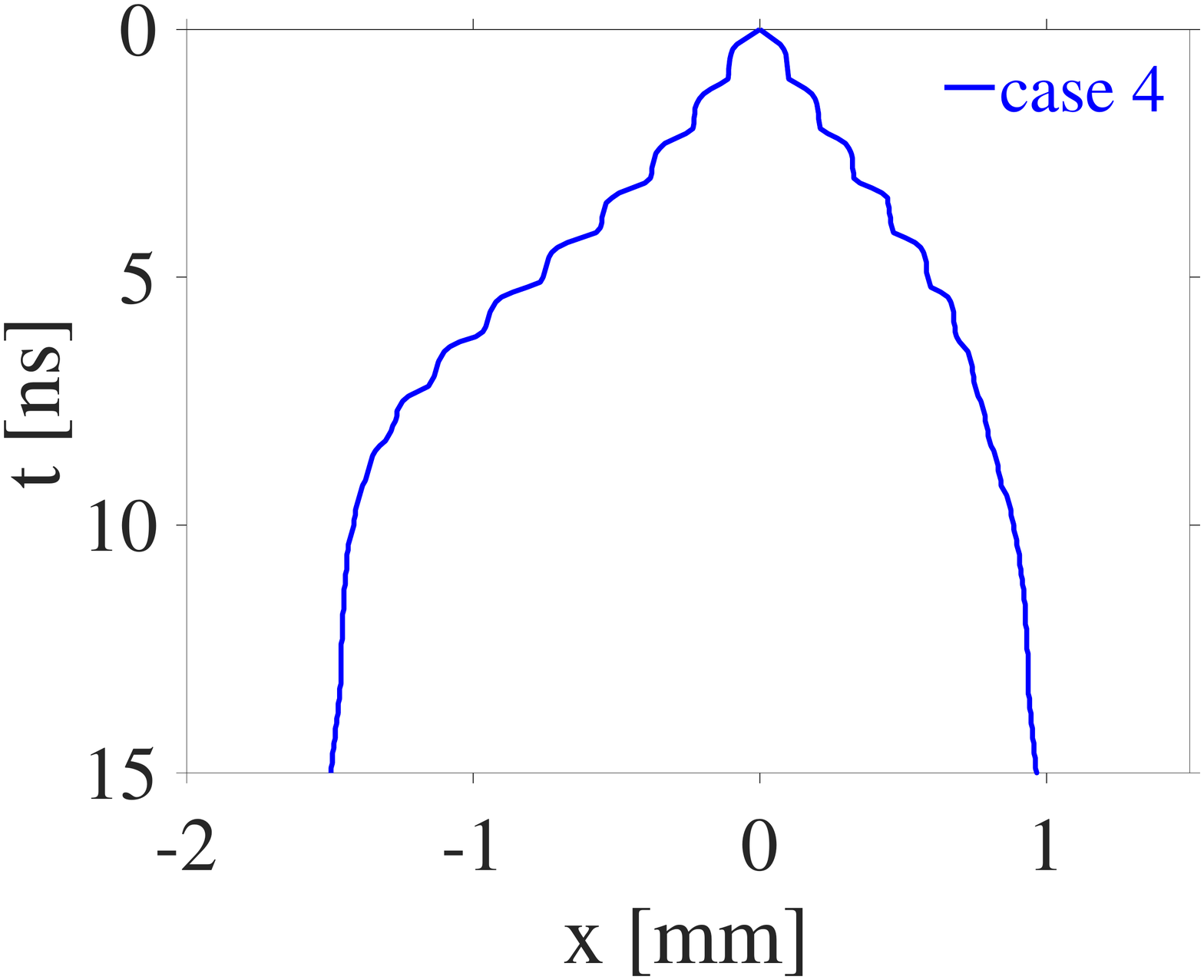}
\hspace{-0.7cm}
\includegraphics[width=0.35\textwidth]{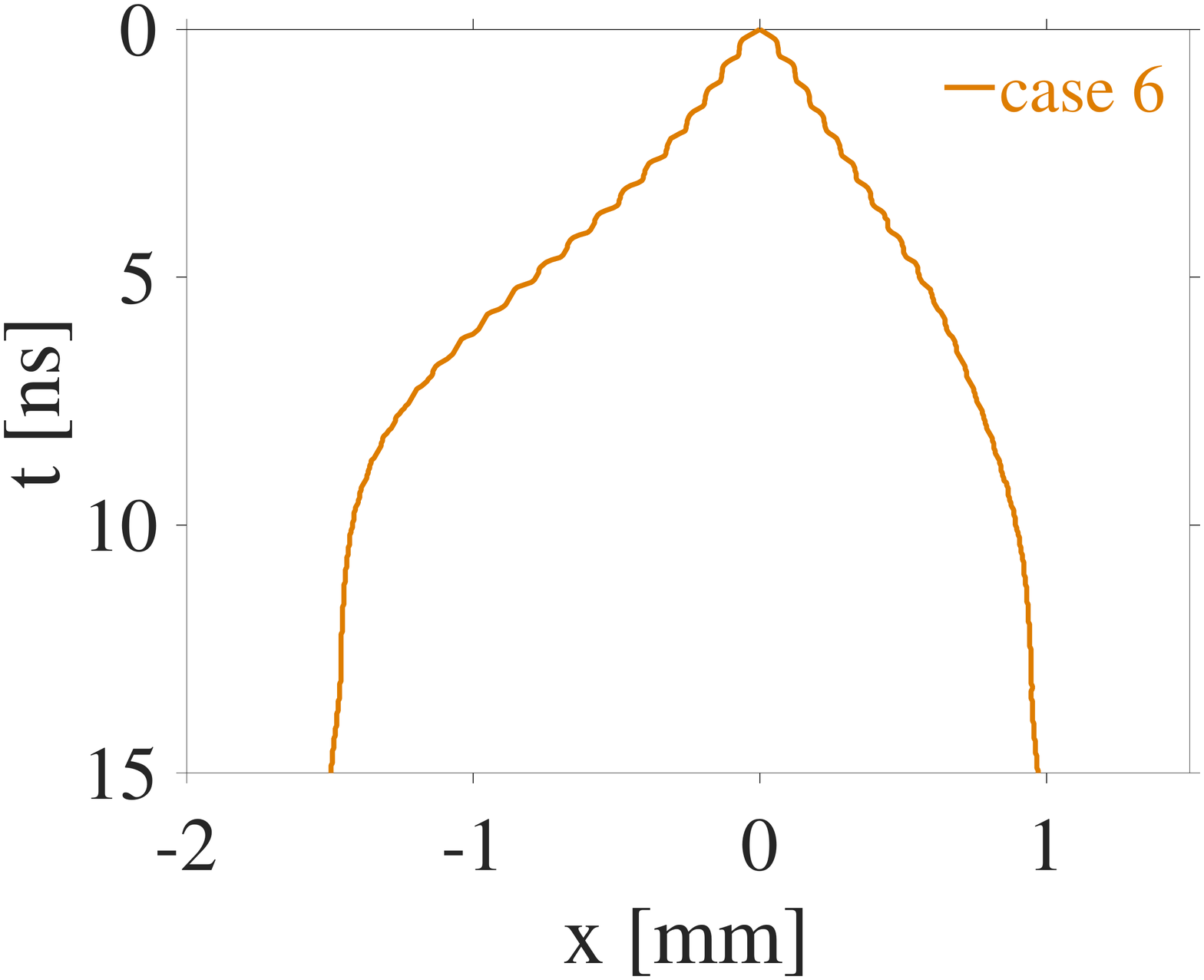}
\caption{Temporal evolution of the maximum number density of of free-electrons (top) and $x$-$t$ diagram of plasma boundary along the optical axis (bottom).}
\label{fig:multi_max_val_analysis2}
\end{center}
\end{figure*}

\begin{figure}[htp]
\begin{center}
\includegraphics[width=0.5\textwidth]{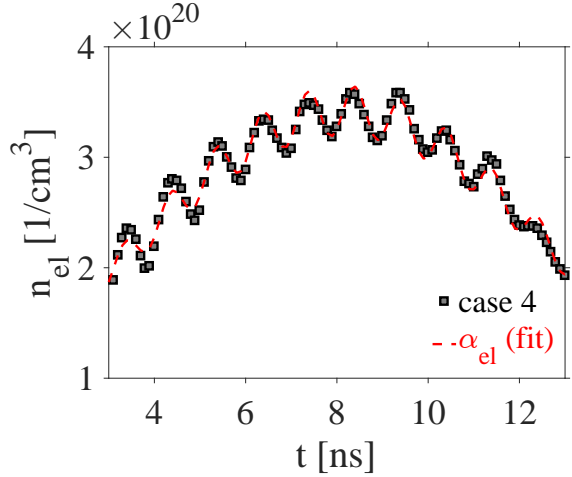}
\vspace*{-.2cm}
\caption{Maximum value of the number density of free-electrons during the multi-mode discharge for case 4 (symbol) compared with amplitude-modulated Gaussian fit \eqnref{eq:wave_fit} (broken line).}
\label{fig:multi_max_case4_fit}
\end{center}
\end{figure}

In order to asses the cause of the observed local in-homogeneities (\emph{i.e.}, quasi-periodic structures) in the plasma kernel, the evolution of the maximum values of the number density of free-electrons and the plasma boundary (using the $x$-$t$ diagram) are monitored for cases 1 through 6 (see Fig. \ref{fig:multi_max_val_analysis2}). It may be seen that the alternation between local maxima and minima, in the maximum number density plots, occurs with a period of $\approx \SI{2}{\nano\second}$ for case 2, $\approx \SI{1}{\nano\second}$ for case 4, and $\approx \SI{0.5}{\nano\second}$ for case 6. Those values are precisely the inverse of the mode-beating frequencies (\emph{i.e.}, periods) of the corresponding cases: $\SI{1/500}{\mega\hertz^{-1}}$ for case~2, $\SI{1/1000}{\mega\hertz^{-1}}$ for case~4, and $\SI{1/2000}{\mega\hertz}^{-1}$ for case 6. To put these observations on a quantitative basis, the predicted maximum number density data for case 4 are compared with an amplitude-modulated Gaussian fit which mimics the time-dependence of the laser beam (see Fig.~\ref{fig:multi_max_case4_fit}):
\begin{widetext}
\begin{equation}\label{eq:wave_fit}
n^{\max}_{\mathrm{e}} =  n_0  \left[ a_1 + a_2 \sin\left(2 \pi a_3 t + a_4\right) \right] \exp\left[-\frac{\left( t - a_5/2\right)^2}{2 a_6^2}\right],  
\end{equation}
\end{widetext}
where $ n_0 =  \SI{5.5e+20}{\centi\meter^{-3}}$, $ a_1 = 0.6219$, $ a_2 = 0.04116$, $ a_3 = \SI{992.2}{\mega\hertz}$, $ a_4 = -0.4493$, $ a_5 = \SI{16.27}{\nano\second}$ and $ a_6 = \SI{4.829}{\nano\second}$. The results of the simulation are in very good agreement with Eq. \ref{eq:wave_fit}. Moreover, the value of the $a_3$ coefficient, which represents the modulation frequency of the signal \eqnref{eq:wave_fit}, is very close to the mode-beating frequency of case 4 (\emph{i.e.}, \SI{1000}{\mega\hertz}). Similar outcomes are found for the other multi-mode cases investigated here. The results in Fig. \ref{fig:multi_max_case4_fit} seem to suggest that the observed quasi-periodic structures are directly linked to the mode-beating frequency, and not induced by hydrodynamic instabilities. In similar fashion of single-mode lasers \cite{alberti_JPhysD}, the spatio-temporal evolution of the plasma and the formation of the kernels is triggered by multiphoton ionization, which creates priming electrons in the  undisturbed gas at the edges of the evolving plasma. Then, the mechanism is sustained by energy deposition via inverse Bremsstrahlung, which counteracts the energy losses due to the cascade ionization. The dynamics of the plasma lobes is guided by both multiphoton ionization and ionization by electron impact, with a strong radial plasma expansion that prevents the plasma kernel from becoming completely opaque to the incoming radiation. 

\section{\label{sec:conc}Conclusions}
This work discussed the application of a physical model, previously developed by the authors, to study the effects of mode-beating on the dynamics of laser-induced gaseous plasmas. The plasma was treated via a fluid approach based on the Navier-Stokes equations. Non-equilibrium effects were accounted for based on a two-temperature model. The propagation and attenuation of the laser beam were modeled by means of the Radiative Transfer equation, which was self-consistently coupled to the flow governing equations. Numerical solutions were obtained based on an Implicit-Explicit (IMEX) finite volume method.  

Applications considered laser-induced breakdown in atmospheric air for nano-second multi-mode pulses, constructed based on amplitude-modulated Gaussian time-profiles. The results of the simulations were first compared with experiments for the sake of validation. An overall good agreement was observed for the absorbed energy, evolution of the plasma boundary and electron number density at the end of the discharge. The calculations were also able to reproduce key features found in experiments: a two-lobed plasma kernel (as in the single-mode configuration) and formation of quasi-periodic structures. It is worth mentioning that this was achieved without seeding the ambient gas with an artificial plasma (as often done in the literature). An analysis of the results showed that the quasi-periodic structures are strongly affected by the mode-beating frequency, as observed also experimentally, and that their occurrence corresponds to the local power peaks of the incident beam. Moreover, the simulations suggest that the laser mode-beating frequency does not appear to seed hydrodynamic instabilities (via interaction between excited and transmitted waves at the plasma surface), but instead directly leads to the periodic kernel structures that correspond to the operating beating-mode. The modulation frequency of the laser determines the local electron density and temperature, with the multi-mode pulses leading to a more elongated plasma, compared to the single-mode case. This depends on the correlation between the temporal peaks/valleys of the laser pulse and the gas-dynamic expansion of the plasma core, that influences the heating rate with the laser radiation being able (or not) to escape more effectively the plasma rear lobe and being absorbed by the front lobe. Future work will focus on characterizing the breakdown mechanism dependence on multi-mode change of phase of laser intensity due to constructive or destructive interference between modes.

\begin{acknowledgments}
This material is based in part upon work supported by the Department of Energy, National Nuclear Security Administration, under Award Number DE-NA0002374. The authors gratefully acknowledge Dr.~J.~H.~Cooley and Dr.~T.~Wada for the useful scientific discussions. 
\end{acknowledgments}

\section*{\label{sec:DataAval}Data Availability}
The data that supports the findings of this study are available within the article and its references.

\appendix

\section{\label{appendixA}The equation of radiative transfer}
In this work, the radiation field is modled with a kinetic approach through the Radiative Transfer Equation (RTE). The labeling kinetic approach stems from Kinetic Theory, and means that radiation is treated as a photon gas described by the one-particle distribution function $f = f(\mathbf{r},\, \mathbf{p},\,t)$, where $\mathbf{r}$ and $\mathbf{p}$ denote position and momentum, respectively, whereas $t$ is time. The monochromatic intensity $I_{\lambda} = I_{\lambda}(\mathbf{r},\,\Omega,\,t)$ is related to the distribution function by \cite{oxenious}:
\begin{equation}\label{eq:I}
I_{\lambda} = \frac{c^2 h^4_{\textsc{p}}}{\lambda^5} f,
\end{equation}
where $c$ is the vacuum speed of light and $h_{\textsc{p}}$ is Planck's constant. The RTE is obtained via the combined use of Eq. (\ref{eq:I}) and the kinetic equation for $f$ (see Ref. \cite{oxenious} for more details). As the above assumptions imply, the RTE may account for emission, absorption and scattering of light within the boundaries of geometric optics. The effects of refraction and polarization, may be also included at the price of a more complicated formulation.  

In alternative to the RTE, one may consider Maxwell's equations. Here dispersive and dissipative effects are contained in the constitutive relations. For instance, the electric displacement (in the frequency domain) for isotropic media may be written:
\begin{equation}\label{eq:Domega}
\mathbf{D}(\omega) = \varepsilon (\omega) \mathbf{E} (\omega),
\end{equation}
where $\varepsilon = \varepsilon (\omega)$ is the medium (complex) dielectric function in the frequency domain. In the time domain, Eq. (\ref{eq:Domega}) translates into a convolution integral. The main advantage of Maxwell's equations compared to an RTE-based approach, is the possibility of including wave phenomena (\emph{e.g.}, diffraction) which cannot be described by geometric optics.

\bibliography{bibliography}

\end{document}